\documentclass[12pt]{article}
\usepackage{xcolor, soul}
\sethlcolor{yellow}
\usepackage{amsmath}
\usepackage{amssymb}
\usepackage{hyperref}
\begin{document}
\vskip 2cm
\begin{center}
{\bf {\Large FLPR Model: (Anti-)Chiral Supervariable Approach to Quantum Symmetries}}\\

\vskip 1cm

{\sf Bhupendra Chauhan $^{(a)}$, $\,$ Tapobroto Bhanja$^{(b)}$}\\\vskip 0.4 cm

{\it $^{(a)}$ Department of Physics, Sunbeam Women's College Varuna,  \\Varanasi, 221 002 (Uttar Pradesh), India}\\
\vskip 0.2cm
{\it $^{(b)}$Department of Physics, University of Science and Technology Meghalaya, Ri-Bhoi - 793101, (Meghalaya), India}\\

\vskip 0.3cm


\vskip 0.1cm

{\small {\sf {E-mails: bchauhan501@gmail.com; tapobroto.bhanja@gmail.com}}}

\end{center}
\date{\today}

\vskip .8cm

\noindent
{\bf Abstract:} We discuss and derive  the  off-shell nilpotent of order two and absolutely anti-commuting
Becchi-Rouet-Stora-Tyutin (BRST), anti-BRST, co-BRST and anti-co-BRST
symmetry transformations for the non-interacting  Friedberg-Lee-Pang-Ren (FLPR) model  in one (0 + 1)-dimension (1D) of spacetime by exploiting the standard techniques of
the (anti-)chiral supervariable approach (ACSA) onto (1, 1)-dimensional super sub-manifold of the general (1, 2)-dimensional supermanifold, where the (anti-)BRST and (anti-)co-BRST invariant restrictions play a crucial role. We provide clear proof of nilpotency and absolute anti-commutativity properties of the (anti-)BRST as well as  
(anti-)co-BRST Noether's conserved charges within the framework of ACSA to BRST formalism, 
where we take only one Grassmannian variable in place of two usual Grassmannian variables   (i.e., fermionic variables). Furthermore, we also demonstrate that the Lagrangian of this non-interacting FLPR model 
  is  (anti-)BRST as well as  (anti-)co-BRST symmetries invariance
within the ambit of the ACSA to BRST approach in (1, 1)-dimensional super sub-manifold.\\

\vskip 1.0 cm



\vskip 0.2cm

\noindent
\section{Introduction}

\vskip 0.2 cm

Gauge theory is a fundamental framework in theoretical physics that describes the dynamics of fields in the realm of particle
physics and quantum field theory [1]. This theory is defined by the presence of first-class constraints, according to Dirac's classification scheme for constraints [2, 3]  where the equations of motion retain their form 
under changes in coordinates. Generally, these transformations facilitate problem-solving by ensuring that the resulting 
solutions remain physically valid and coherent.
{In gauge theory, fields are associated with {certain} symmetries called gauge symmetries (symmetry transformations).
These symmetries imply that the physics of the system does not depend on the choice of a specific reference
frame or ``gauge"  used to describe it.} Gauge theories are essential in understanding the behaviour of the 
fundamental forces of nature such as electromagnetism, the weak nuclear force, and the strong nuclear force. For instance, the standard model
of particle physics, which describes the electromagnetic, weak, and strong interactions, is based on gauge theories.
This theory has profound implications not only for particle physics but also for other areas of physics, 
such as condensed matter physics and cosmology, where gauge theories play important roles in understanding
the behaviour of complex systems {and the early Universe.}

\vskip .2cm

The covariant quantization of a gauge-field system has a long history starting from
the famous works of Feynman [4], Faddeev and Popov [5], and DeWitt [6].
The BRST (Becchi, Rouet, Stora, and Tyutin) formalism [7-10] provides a powerful framework for quantizing gauge theories
and understanding their properties at the quantum level. The BRST formalism is often used for gauge theories to handle redundant degrees of freedom and ensure consistent quantization of constrained systems. It ensures the consistency and unitarity of the theory 
while preserving gauge invariance.  In this approach,
the infinitesimal local gauge parameter is replaced by ghost and anti-ghost fields to maintain the unitarity of the theory. At the quantum level, 
this leads to the existence of two global supersymmetric-type BRST ($s_b$) and anti-BRST ($s_{ab}$) symmetry transformations. These transformations
have two key mathematical properties: first, they are nilpotent of order two (i.e., $s_b^2 = 0$ and $s_{ab}^2 = 0$), and second, they exhibit absolute anticommutativity (i.e., $s_b s_{ab} + s_{ab} s_b = 0$). The nilpotency indicates that the BRST and anti-BRST symmetry transformations are fermionic,
while the anticommutativity shows that they are linearly independent. A distinctive feature of BRST-quantized (non-)Abelian gauge theories is the presence
of these Curci-Ferrari (CF) {or CF-type restrictions} at the quantum level. The CF condition is an (anti-)BRST invariant quantity, signifying its role as a physical 
condition in quantum theory. Furthermore, advanced covariant quantization methods for general gauge theories are based on either the BRST symmetry 
principle, as realized in the quantization scheme developed by Batalin and Vilkovisky, or the extended BRST symmetry principle, as utilized in the
quantization method proposed by Batalin, Lavrov, and Tyutin [11-13].

\vskip .2cm

The Friedberg-Lee-Pang-Ren (FLPR) model  provides a theoretical framework for understanding the dynamics of a single non-relativistic
particle in the presence of a general two-dimensional rotationally invariant potential. This model has been discussed in detail in the paper
for phase transition and symmetry breaking in the FLPR model [14, 15]. The FLPR model, introduced by Friedberg, Lee, Pang, and Ren, is a solvable gauge model of a single non-relativistic particle with a unit mass that 
exhibits characteristics of Gribov ambiguity [16] and
this model serves as a valuable tool for investigating the 
behaviour of particles subjected to potentials dependent on the square of their position coordinates, offering insights into various
phenomena in quantum mechanics. The FLPR model has also been analyzed using a 
gauge independent method to abstract the reduced physical space, with complications related to gauge fixing being explored [17]. Moreover,
the advantages of employing a physical projector in the quantization of gauge-invariant systems have been investigated in the context of
the FLPR model [18].
{The Friedberg-Lee-Pang-Ren (FLPR) model, analyzed within the framework of the BRST  formalism, provides a robust approach to understanding
its quantum properties and symmetries [19].} 
 By incorporating auxiliary
fields and ghost degrees of freedom, the BRST approach enables the construction of BRST invariant Lagrangians for the FLPR model. This formalism
not only identifies the gauge symmetries associated with rotational invariance but also provides a systematic method for quantizing the theory 
consistently. Additionally, the BRST formalism elucidates the structure of the FLPR model's Hilbert space and the role of gauge-fixing terms 
in maintaining gauge symmetry. Through this approach, researchers deepen their understanding of the FLPR model's quantum behaviour, offering 
insights into phenomena such as phase transitions and symmetry breaking.  A comprehensive BRST analysis has been conducted by summing over all Gribov-type copies [20]. Recently, the FLPR model has been discussed within the BRST formalism
for the various theoretical points of view [21-24]. 

\vskip .2cm

The traditional superfield approach (USFA) to BRST formalism [25-29] leverages the horizontality condition (HC) to derive off-shell nilpotent 
(anti-)BRST symmetry transformations for gauge, ghost, and anti-ghost fields, incorporating full super expansions of the superfield involving 
two Grassmann variables $(\vartheta, \bar\vartheta)$. However, this approach does not address the matter fields in interacting theories. To 
remedy this, an extended version known as the augmented version of the superfield approach (AVSA) was developed, which uses both HC and gauge
invariant restriction(s) (GIR) to derive (anti-)BRST symmetries for matter fields [30-32]. This approach has been extensively applied to general
gauge theories, offering a geometric interpretation of BRST quantization [33-35].  Building on this, we applied the newly introduced (anti-)chiral 
superfield approach (ACSA) to derive a comprehensive set of (anti-)BRST symmetries by utilizing (anti-)chiral super expansions with a single 
Grassmann variable [36-47]. We also combined ACSA with the modified Bonora-Tonin superfield approach (MBTSA) to derive complete (anti-)BRST 
symmetry transformations for various reparameterization invariant models [48-50]. 
The novelty of our present work depends on the facts that: (i) absolute anticommutativity and nilpotency of the (anti-)BRST and (anti-)co-BRST symmetry transformations are derived for the FLPR model by using only the (anti-)chiral supervatiables which are defined on the (1,1)-dimensional super-sub-manifold with only one Grassmannian variable (either $\vartheta$ or $\bar\vartheta$) instead of two variables $(\vartheta, \bar\vartheta)$ of the original (1,2)-dimensional supermanifold containing bothe the supervariables. This is interesting and provides a simpler way of deriving these fermionic symmetries and highlight the richness and strength of the ACSA approach making the calculations simpler, (ii) we have not used the (dual-)horizontality conditions in our derivations anywhere. We have instead explored the strength of the (anti-)BRST and (anti-)co-BRST restrictions defined on the (1,1)-dimensional super-sub-manifold instead of the regular (1, 2)-dimensional supermanifold which is essential for the (dual-)horizontality conditions, in order to derive these conditions, which again depicts the strength and richness of this approach.

In our present endeavor, we derive the nilpotent (anti-)BRST and (anti-)co-BRST symmetry transformations of the theory using the standard techniques of ACSA   where we use the (anti-)chiral super expansions of the variables with only one Grassmannian variable (either $\vartheta$ or $\bar\vartheta$) instead of two variables $(\vartheta, \bar\vartheta)$. Therefore, ACSA is a simpler version in comparison to full super expansions of variables with two Grassmannian variables because ACSA simplifies our mathematical complexity. In other words, we break general (1, 2)-dimensional supermanifold into two simpler versions of (1, 1)-dimensional super sub-manifolds within the framework of ACSA to BRST formalism.

\vskip .2cm  

The order of the various sections of our current paper is as follows. In Sec. 2, we discuss
the (anti-)BRST  and (anti-)co-BRST symmetry transformations for the FLPR model
and deduce the Nilpotent conserved charges. Our Sec. 3 deals with the ACSA to BRST formalism where
 we derive the (anti-)BRST  symmetry transformations. Sec. 4  is devoted
 to the derivation of (anti-)co-BRST symmetry transformations by using the ACSA to BRST formalism where the super expansions of 
(anti-)chiral supervariables are utilized fruitfully. In Sec. 5, we express the conserved (anti-)BRST  and (anti-)co-BRST  charges on
the (1, 1)-dimensional super sub-manifolds [of the general  (1, 2)-dimensional supermanifold]
 on which our theory is generalized and provides proof of nilpotency and absolute
anti-commutativity properties of the (anti-)BRST along with  (anti-)co-BRST charges within the ambit of ACSA
to BRST formalism. Section 6 delves into the (anti-)BRST and (anti-)co-BRST invariances of the Lagrangian within the ACSA framework. Finally, in Section 7, we summarize our key findings and results while also suggesting potential avenues for future research.\\


\noindent
\section{Preliminaries: Gauge Symmetries and its Generator and (Anti-)BRST Symmetries for the  FLPR Model}

We start with the primary Lagrangian ($L_f$) of the FLPR model, representing the motion of a single non-relativistic particle with a mass of unity. 
This particle moves within a spatial 2D environment and experiences the effects of a rotational potential that is invariant regardless of the 
particle's orientation. This potential is described by the function $U (x^2 + y^2)$, which depends solely on the radial distance from the origin.
The Lagrangian expressed in the Cartesian coordinate system takes the following form:
\begin{eqnarray}
L_{f} = p_x \, \dot x + p_y \, \dot y + p_z \, \dot z - \frac{1}{2}\, \big (p_x^2 + p_y^2 + p_z^2 \big ) 
-\zeta\, \bigl [g (x\,p_y - y\, p_x) + p_z \bigr ] - U(x^2 + y^2),
\end{eqnarray}
where $\dot x, \dot y, \dot z$ are the time derivative of the Cartesian coordinates $(x, y, z)$ (i.e. generalized velocities)
which are expressed as $\dot x = (dx/dt), \dot y = (dy/dt), \dot z = (dz/dt)$ and $(p_x, p_y, p_z)$ are the canonical conjugate
momenta corresponding to the coordinates $(x, y, z)$ constrained through the Lagrange multiplier variable
$\zeta (t)$ which satisfies the following relationship: $ g \,(x\,p_y - y\, p_x) + p_z \approx 0$ where $g$ is 
the real positive (i.e. $g >0$) coupling constant.

\vskip 0.2cm

It is straightforward to note that the canonical conjugate momentum $p_\zeta$  w.r.t. $\zeta$ implies  
\begin{eqnarray}
&&p_\zeta = \frac {\partial L_f }{ \partial \dot \zeta} \implies p_\zeta \approx 0, \nonumber\\
&&\frac{d}{dt}\, \Bigl (\frac{\partial L_{f}} {\partial \dot \zeta} \Bigr ) = \frac{\partial L_{f}} {\partial \zeta} \quad
\Longrightarrow \quad \dot p_\zeta = -\,\bigl [g\, (x\,p_y - y\, p_x) + p_z \bigr ] \approx 0,
\end{eqnarray}
primary and secondary constraints on the theory.
 There are no further constraints on our theory because, 
 at this stage itself, it can be seen that  both the above
constraints commute with each other. Therefore, the primary Lagrangian (1) is endowed with first-class constraints
in the terminology of Dirac's prescription for the classification 
scheme of constraints [2, 3].

\vskip 0.2cm

The existence of the first-class constraints [e.g. $p_\zeta \approx 0, \, g\, (x\,p_y - y\, p_x) + p_z \approx 0$] on our theory
ensures that our theory is a gauge theory which obeys the  
following local, continuous and infinitesimal {\it classical} gauge symmetry transformations ($\delta_g$) 
 \begin{eqnarray}
 \delta_g x &=& -\, g\, y \,\alpha, \quad \delta_g y = g \,x \,\alpha, \quad \delta_g z = \alpha, \quad \delta_g p_x = -\, g\, p_y\, \alpha, \nonumber\\
\delta_g p_y &=& g\,p_x \,\alpha, \quad \delta_g p_z = 0, 
\quad \delta_g p_\zeta  = 0, \quad \delta_g \zeta  = \dot \alpha, \quad (\delta_g L_f = 0),    
\end{eqnarray}
are generated by the generator ($G$) that can be precisely
expressed in terms of the above first-class constraints of our theory (see, e.g. [2, 3] for details)
\begin{eqnarray}
G = \dot \alpha\, p_\zeta + \alpha\, [g\, (x\, p_y - y\, p_x) + p_z].
\end{eqnarray}
The parameter \( \alpha(t) \) characterizes an infinitesimally small gauge symmetry transformation. Utilizing the generator 
outlined above allows us to derive the classical gauge symmetry transformations (\( \delta_g \)). This process is evident 
when we employ the standard connection between the infinitesimal continuous gauge symmetry transformation (\( \delta_g \))
for any variable \( \phi(t) \) within our theory, as specified in Equation (1), and the generator \( G \) as defined in Equation (4).
\begin{eqnarray}
\delta_g \, \phi (t) = -i\, [ \phi (t), \;G], \qquad \phi = x, y, z, \zeta, p_x, p_y, p_z, p_\zeta.
\end{eqnarray}
 It is given the standard non-zero canonical commutators (in natural units: \( \hbar = c = 1 \)), we have the following commutators for our theory:
\begin{eqnarray}
 [x, \,p_x] = i, \,\,[y,\, p_y] = i,\,\,[z, \, p_z] = i,\,\,[\zeta, \, p_\zeta] = i.
 \end{eqnarray}
All other canonical commutators of the various fields of Lagrangian ($L$) are defined to be zero according to the rules set by the canonical quantization scheme.

\vskip 0.2cm

We can elevate the classical infinitesimal gauge symmetry transformations (\( \delta_g \)) to their quantum counterparts:
infinitesimal, continuous, and off-shell nilpotent (\( s_{(a)b}^2 = 0 \)), absolutely anticommuting (\(s_b \, s_{ab} + s_{ab}\, s_b = 0\)) (anti-) BRST symmetry transformations (\( s_{(a)b} \)) as follows:
\begin{eqnarray}
 &&s_{ab} \, x = -\, g\, y \,\bar c, \quad s_{ab}\, y = g \,x \,\bar c, \quad s_{ab}\, z = \bar c, \quad s_{ab}\, \zeta  = \dot {\bar c}, \quad s_{ab}\, p_x = -\, g\, p_y\, \bar c, \nonumber\\
&&s_{ab}\, p_y =  g \,p_x \,\bar c, \quad s_{ab}\, p_z = 0, \quad s_{ab}\, p_\zeta  = 0,  \quad s_{ab} \,\bar c = 0, \quad s_{ab}\,c = - i\, b, \quad s_{ab}\,b = 0, \nonumber\\
 &&s_{b} \, x = -\, g\, y \,c, \quad s_{b}\, y = g \,x \, c, \quad s_{b}\, z =  c, \quad s_{b}\, \zeta  = \dot  c, \quad s_{b}\, p_x = -\, g\, p_y\,  c, \nonumber\\
&&s_{b}\, p_y =  g \,p_x \, c, \quad s_{b}\, p_z = 0, \quad s_{b}\, p_\zeta  = 0,  \quad s_{b} \, c = 0, \quad s_{b}\,\bar c =  i\, b, \quad s_{b}\,b = 0.
\end{eqnarray}
These transformations are the symmetry transformations for the generalized version of the first-order
classical Lagrangian (\( L_f \)) to its quantum (anti-)BRST invariant Lagrangian (\( L \)).

\vskip 0.2cm

The Lagrangian \( L\) that incorporates the 't Hooft-like gauge-fixing term and the fermionic Faddeev-Popov 
ghost terms are given by:
\begin{eqnarray}
L &=&  p_x \, \dot x + p_y \, \dot y + p_z \, \dot z - \frac{1}{2}\, \big (p_x^2 + p_y^2 + p_z^2 \big ) -\zeta\, \bigl [g (x\,p_y - y\, p_x) + p_z \bigr ] \nonumber\\
& - & U(x^2 + y^2)  + b \,\bigl (\dot \zeta - z \bigr ) + \frac{1}{2}\, b^2 - i \, \dot {\bar c}\, \dot c -\, i\,\bar c\, {c}. 
\end{eqnarray}
We observe the following transformations of \( L \) under the (anti-)BRST symmetries:
\begin{eqnarray}
 s_{ab} \, L  = \frac{d}{dt} \, \Big[b\, \dot {\bar c} \Big], ~\qquad\qquad  s_{b} \, L = \frac{d}{dt} \, \Big [b\,  \dot c \Big], 
\end{eqnarray}
which shows that under the (anti-)BRST symmetry transformations, Lagrangian transforms to the total time derivative, which establishes the (anti-)BRST invariance of the action integral corresponding to the Lagrangian $( L)$ of the present FLPR model.

\vskip 0.2cm

We end this section with the following concluding remarks (i) the nilpotency property shows that the (anti-)BRST symmetry transformations are fermionic,  (ii) these transformations are absolutely anticommuting, establishing the linear independence of the BRST and anti-BRST symmetry transformations,   (iii) the first-order Lagrangian \( L_f \) is also (anti-)BRST invariant and {the total kinetic terms of the theory remain invariant under the (anti-)BRST symmetry transformations,}  (iv) the gauge variable is identified through the gauge and the (anti-)BRST symmetry transformations.
The infinitesimal (anti-)BRST symmetry transformations are generated by the conserved (anti-)BRST charges:
\begin{eqnarray}
{\cal Q}_{ab} &= \bigl [g \, (x\, p_y - y \, p_x) + p_z \bigr ]\, \bar c + b \, \dot {\bar c} \equiv b\, \dot {\bar c} - \dot b \, \bar c, \nonumber\\
{\cal Q}_{b} &= \bigl [g \, (x\, p_y - y \, p_x) + p_z \bigr ]\,  c + b \, \dot {c} \equiv b\, \dot {c} - \dot b \, c. 
\end{eqnarray}
The conservation laws for the above (anti-)BRST charges can be proven using the equations of motion, derived from the Lagrangian $(L)$ for the relevant variables.

\section{(Anti-)co-BRST Symmetry Transformations}

The (anti-)BRST invariant Lagrangian \(L\) [cf. Eq. (8)], in addition to the (anti-)BRST symmetry transformations, also obeys another set of off-shell nilpotent (i.e. \(s_{(a)d}^2 = 0\)) and absolutely anticommuting (i.e. \(s_d \, s_{ad} + s_{ad}\, s_d = 0\)) (anti-)dual-BRST symmetry transformations. In literature, these fermionic symmetries are also christened as the (anti-)co-BRST symmetry transformations under which the gauge-fixing terms remain invariant. For our theory under consideration (described by the Lagrangian \(L\)), these infinitesimal, continuous, off-shell nilpotent {and absolutely anticommuting} (anti-)co-BRST symmetry transformations \(s_{(a)d}\) are given by:
\begin{eqnarray}
s_{ad} \, x &= -\, g\, y \,\dot {c}, \quad s_{ad}\, y = g \,x \,\dot c, \quad s_{ad}\, z = \dot c, \quad s_{ad}\, \zeta  = c, \quad s_{ad}\, p_x = -\, g\, p_y\, \dot c, \quad s_{ad}\,c = 0, \nonumber\\
s_{ad}\, p_y &=  g \,p_x \,\dot c, \quad s_{ad}\, p_z = 0, \quad s_{ad}\, p_\zeta  = 0,  \quad s_{ad} \,\bar c = i\, \bigl [g (x\, p_y - y\, p_x) + p_z \bigr ], \quad s_{ad}\,b = 0, \nonumber\\
s_{d} \, x &= -\, g\, y \,\dot {\bar c}, \quad s_{d}\, y = g \,x \,\dot {\bar c} , \quad s_{d}\, z = \dot {\bar  c}, \quad s_{d}\, \zeta  = \bar c, \quad s_{d}\, p_x = -\, g\, p_y\, \dot {\bar c}, \quad s_{d}\,b = 0,\nonumber\\
s_{d}\, p_y &=  g \,p_x \, \dot {\bar c}, \quad s_{d}\, p_z = 0, \quad s_{d}\, p_\zeta  = 0,  \quad s_{d} \, \bar c = 0, \quad s_{d}\, c = - \, i\, \bigl [g (x\, p_y - y\, p_x) + p_z \bigr ].   
\end{eqnarray}
It's evident from these transformations that \(s_{(a)d}\, [(\dot \zeta - z)]  = 0\) and \(s_{(a)d}\, b = 0\), ensuring the total gauge-fixing terms remain invariant.

\vskip 0.2cm

Under the (anti-)co-BRST symmetry transformations, it's straightforward to verify that the Lagrangian \(L\) transforms to total time derivatives:
\begin{eqnarray}
s_{ad}\, L &= \frac{d}{dt}\, \Bigl [ \{g \, (x\, p_y - y \, p_x) + p_z \}\, \dot c \Bigr ], \quad s_{d}\, L = \frac{d}{dt}\, \Bigl [ \{g \, (x\, p_y - y \, p_x) + p_z \}\, \dot {\bar c} \Bigr ],
\end{eqnarray}
demonstrating that the action integral \(S = \int dt\, L\) remains invariant since all physical and ghost variables vanish as \( t \to \pm \, \infty \). According to Noether's theorem, the invariance of the action integral leads to the derivation of the (anti-)co-BRST charges \({\cal Q}_{(a)d}\):
\begin{eqnarray}
{\cal Q}_{ad} &= \bigl [g\, (x\, p_y - y \, p_x) + p_z \bigr ] \, \dot c + b\, c \equiv b\, c - \dot b\, \dot c, \nonumber\\
{\cal Q}_{d} &= \bigl [g\, (x\, p_y - y \, p_x) + p_z \bigr ] \, \dot {\bar c} + b\, \bar c \equiv b\, {\bar c} - \dot b\, \dot {\bar c}.
\end{eqnarray}
The  conservation of charges can be proven using the relevant equations of motion derived from \(L\), such as:
\begin{eqnarray}
&&\dot x = p_x - g\,\zeta\,y, \quad \dot y = p_x + g\,\zeta\,x, \quad \dot z = p_z 
+ \zeta, \quad \dot p_z = -\, b, \nonumber\\
&&\dot p_x = - g\,\zeta\,p_y -\, 2\,x\, U^\prime, \quad \dot p_y = g\, \zeta\,p_x  -\, 2\,y\, U^\prime, \quad 
\dot b = -\, \bigl [g\, (x\, p_y - y \, p_x) + p_z \bigr ].
\end{eqnarray}
The derivative {condition} \(\ddot b = b\), derived using EoMs, plays a crucial role in proving the conservation of the (anti-)co-BRST charges (13), along with the 
equations of motion for the ghost variables with the additional help from the EL-EoMs: $\ddot c = c, \; \ddot {\bar c} = {\bar c}$.

\vskip 0.2cm

We conclude this section with several important observations. Firstly, it is  essential to note that while the total kinetic terms 
of the Lagrangian $(8)$ for our FLPR model remain invariant under the (anti-)BRST symmetry transformations, the total gauge-fixing
terms remain invariant under the (anti-)co-BRST symmetry transformations. This distinction arises from the operations of exterior
and co-exterior derivatives in differential geometry, which are foundational in gauge theories.
Secondly, the off-shell nilpotency and absolute anti-commutativity properties of the (anti-)co-BRST transformations hold, as evidenced, 
by $s_{(a)d} \, \bigl [g (x\, p_y - y\, p_x) + p_z \bigr ] = 0$.
{Thirdly, it is interesting to highlight that the first-class constraints of our theory are invariant under the infinitesimal
gauge symmetry transformations, nilpotent (anti-) BRST symmetry transformations, and nilpotent (anti-)co-BRST symmetry transformations. }
These {constraints} are crucial physical restrictions in our theory, both classically and quantum-mechanically.
Lastly, it is important to recognize that the (anti-) BRST and (anti-)co-BRST symmetry transformations are fermionic, transforming
bosonic variables into fermionic ones and vice versa.

\section{Nilpotent Quantum (Anti-)BRST Symmetry Transformations: (Anti-)Chiral Supervariable Approach}

\vskip 0.2cm

In this section, we discuss the step-by-step derivation of the nilpotent BRST  and anti-BRST symmetry transformations [see Eq. (7)] by employing the (anti-)chiral supervariable approach (ACSA) to BRST formalism [36-42]. This involves utilizing the (anti-)chiral super expansions of supervariables taken from Lagrangian $L$ and (anti-)BRST invariant restrictions. To find the BRST symmetry transformations of the ordinary variables of the Lagrangian, first of all,  we extend the ordinary variables present in the Lagrangian of FLPR model (8) onto a (1, 1)-dimensional anti-chiral super sub-manifold, which is a subset of the general  (1, 2)-dimensional supermanifold, as follows:
\begin{eqnarray}
&&x (t) \quad \; \longrightarrow \quad X (t, \bar\vartheta) \,\;= \; x (t) + \bar\vartheta\,b_1 (t),\nonumber\\
&&y (t) \quad \;\longrightarrow \quad Y (t, \bar\vartheta) \;\;= \; y (t) + \bar\vartheta\,b_2 (t),\nonumber\\
&&z (t) \quad \;\longrightarrow \quad Z (t, \bar\vartheta) \;\,\;= \; z (t) + \bar\vartheta\,b_3 (t),\nonumber\\
&&p_x (t)\quad  \longrightarrow \quad P_x (t, \bar\vartheta) \;= \; p_x (t) + \bar\vartheta\,b_4 (t),\nonumber\\
&&p_y (t) \quad \longrightarrow \quad P_y (t, \bar\vartheta) \;= \; p_y (t) + \bar\vartheta\,b_5 (t),\nonumber\\
&&p_z (t)  \quad\longrightarrow \quad P_z (t, \bar\vartheta) \;= \; p_z (t) + \bar\vartheta\,b_6 (t),\nonumber\\
&&\zeta (t) \,\;\quad \longrightarrow \quad \Xi (t, \bar\vartheta) \;\,\;= \; \zeta (t) + \bar\vartheta\,b_7 (t),\nonumber\\
&&p_\zeta (t) \quad \longrightarrow \quad P_\zeta (t, \bar\vartheta) \,\;= \; p_\zeta (t) + \bar\vartheta\,b_8 (t),\nonumber\\
&&b (t) \quad \;\;\longrightarrow \quad B (t, \bar\vartheta) \,\;= \; b (t) + \bar\vartheta\,b_9 (t),\nonumber\\
&&c (t)  \;\;\quad\longrightarrow \quad F (t, \bar\vartheta) \,\;= \; c (t) + \bar\vartheta\,f_1 (t),\nonumber\\
&&\bar c (t) \;\quad \;\longrightarrow \quad  \bar F (t, \bar\vartheta) \,\;= \; \bar c (t) + \bar\vartheta\,f_2 (t).
\end{eqnarray}
Here $b_1, b_2, b_3, b_4, b_5, b_6, b_7, b_8, b_9$ are the {bosonic}  secondary variables while $f_1$ and  $f_2$ are the {fermionic} secondary variables, attributed to the  fermionic nature of $\bar\vartheta$. The precise values of these derived variables are determined in terms of the auxiliary and basic variables present in the BRST invariant Lagrangian (8) by using the BRST invariant conditions and restrictions.

\vskip 0.2cm

According to the basic principles of ACSA, the BRST invariant quantities must remain independent of the Grassmannian variable ($\bar\vartheta$) when they are generalized onto the (1, 1)-dimensional anti-chiral super sub-manifold. The BRST invariant quantities are the specific combinations of the variables present in Lagrangian (8), given as follows:
\begin{eqnarray}
&& s_b(c, b, p_z, p_\xi) = 0, \quad s_{b} (x\, c) = 0,\quad s_{b} (y\, c) = 0, \quad s_b (x^2 + y^2) = 0, \quad s_b (p_x\, c) = 0, \quad  \nonumber\\
&&s_b (p_y\, c) = 0,\quad s_b (p_x^2 + p_y^2) = 0,\quad s_b (z\,c) = 0, \quad s_b (\zeta\, \dot c) = 0,\nonumber\\
&& s_{b} (\dot\zeta  - z) = 0, \quad s_{b} (\dot b \, \zeta + i\, \dot {\bar c}\,\dot c) = 0,\quad s_{b} (b\,z + i\,\bar c\, c) = 0.
\end{eqnarray}
We generalize the above BRST invariant restrictions onto the (1, 1)-dimensional anti-chiral super sub-manifolds 
(of the suitably chosen most common (1, 2)-dimensional super-manifold):
\begin{eqnarray} 
&& F (t, \bar\vartheta)  = c(t), \; {B} (t, \bar\vartheta)  = b (t), \;  P_z  (t, \bar\vartheta)  = p_z  (t),\;
 P_\zeta  (t, \bar\vartheta)  = p_\zeta  (t),\quad X (t, \bar\vartheta)\, F (t, \bar\vartheta)  =\nonumber\\
 &&  x (t) \, C(t), \quad 
 Y (t, \bar\vartheta)\, F (t, \bar\vartheta)  = y (t) \, c(t),\quad X^2 (t, \bar\vartheta) + Y^2 (t, \bar\vartheta)  = x^2 (t) + y^2 (t),\nonumber\\
 && P_x (t, \bar\vartheta)\, F (t, \bar\vartheta) = p_x (t), \quad P_y (t, \bar\vartheta)\, F (t, \bar\vartheta) = p_y (t)\,  c(t), \quad P_x^2 (t, \bar\vartheta) + P_y^2 (t, \bar\vartheta)  = \nonumber\\
 && p_x^2 (t) + p_y^2 (t), \quad {Z} (t, \bar\vartheta) \, F (t, \bar\vartheta)  = z (t)\, c(t), \quad \Xi (t, \bar\vartheta) \, \dot F (t, \bar\vartheta)  
= \zeta (t)\, \dot c(t),\nonumber\\
&&\dot\Xi (t, \bar\vartheta)  - {Z} (t, \bar\vartheta)  = \dot \zeta  (t) - z (t), \quad {\dot {{B}}} (t, \bar\vartheta)  \,\Xi (t, \bar\vartheta)  +  i\,\dot{\bar F} (t, \bar\vartheta) \, \dot F (t, \bar\vartheta) 
=  \nonumber\\
&&\dot  b (t) \, \xi (t) + i\, \dot {\bar c} (t)\,\dot c (t), \quad { {B}} (t, \bar\vartheta)  \,Z  (t, \bar\vartheta)  + i\,{\bar F} (t, \bar\vartheta) \,  F (t, \bar\vartheta) 
=  {b} (t) \, z  (t) + i\,  {\bar c} (t)\, c (t). 
\end{eqnarray}
The above restrictions lead to the derivation of the secondary variables in terms of the basic and auxiliary variables.
To determine the value of these variables, we perform step-by-step explicit calculations. For this purpose, first of all, we use the generalization of the trivial BRST invariant restrictions given in the first line of Eq. (17) as: 
\begin{eqnarray} 
&&  F (t, \bar\vartheta)  = c(t), \Longrightarrow f_1 = 0, 
\qquad \;\;\; \, P_z  (t, \bar\vartheta)  = p_z  (t) \Longrightarrow b_6 = 0, \nonumber\\
&& P_\zeta  (t, \bar\vartheta)  = p_\zeta  (t), \; \;\Longrightarrow b_8 = 0, \qquad\;\, { B} (t, \bar\vartheta)  = b (t) \Longrightarrow b_9 = 0.
\end{eqnarray} 
After substituting the above value of derived variables from (18) to (15), we get the following expressions 
for the   anti-chiral   supervariables, namely;
\begin{eqnarray}
&& c (t) \;\;\longrightarrow \; F ^{(b)} (t, \bar\vartheta)  = c(t) + \bar\vartheta\,(0) \; \equiv c(t) + \bar\vartheta\,[s_b\, c (t)] ,\nonumber\\
&& p_z (t) \,\longrightarrow \; P_z  ^{(b)} (t, \bar\vartheta)  = p_z (t) + \bar\vartheta\, (0) \equiv p_z (t)
 + \bar\vartheta\, [s_b\, p_z (t)], \nonumber\\
&& p_\xi  (t)\; \longrightarrow \; P_\xi  ^{(b)} (t, \bar\vartheta)  =  p_\xi  (t) + \bar\vartheta\, (0)
 \equiv p_\xi  (t) + \bar\vartheta\, [s_b\, p_\xi  (t)], \nonumber\\ 
&&  {b} (t) \; \;\,\longrightarrow \; {{B}} ^{(b)} (t, \bar\vartheta)  =   b(t) + \bar\vartheta\,(0)\; \equiv  b (t) + \bar\vartheta\,[s_b\, {b} (t)], 
\end{eqnarray}
where the superscript $(b)$ on the anti-chiral supervariables denotes that these supervariables have been obtained after the use of BRST invariant quantities. The coefficients of the Grassmannian variable $\bar\vartheta$ are the BRST symmetry transformations.

\vskip 0.2cm

Now, we consider other non-trivial BRST invariant restrictions, such as $s_{b} (x\,  c) = 0$, and  $s_{b} (y\, c) = 0$ and generalize them onto a $(1, 1)$-dimensional super sub-manifold, we arrive at:
\begin{align} 
 X (t, \bar\vartheta)\, F (t, \bar\vartheta)  = x (t) \, c(t) \Longrightarrow b_1 (t) \propto c (t) \quad \Longrightarrow \quad b_1 (t)  = \kappa_1 \, c (t),  \nonumber\\
 Y (t, \bar\vartheta)\, F (t, \bar\vartheta)  = y (t)\, c(t) \Longrightarrow b_2 (t) \propto c (t) \quad \Longrightarrow \quad b_2 (t)  = \kappa_2 \, c (t). 
\end{align} 
 To determine the  values of $\kappa_1$ and $\kappa_2$, we employ the generalization of the BRST invariant restriction $s_b (x^2 + y^2) = 0$ as:
\begin{align} 
 X (t, \bar\vartheta)\, X (t, \bar\vartheta)  + Y (t, \bar\vartheta)\, Y (t, \bar\vartheta)  = x^2 (t) + y^2 (t) \quad \Longrightarrow \quad b_1 (t)\, x(t) + b_2 (t)\, y(t)  = 0.
\end{align} 
After substituting the values of $b_1$ and $b_2$ from Eq. (19), we obtain the relation: $\kappa_1\,c\,x + \kappa_2\,c\,y = 0$. This relation is valid for the two combinations of values: (i) $\kappa_1 = -\,g\,y, \; \kappa_2 = g\,x$ and   (ii) $\kappa_1 = -\,g\,y, \; \kappa_2 = g\,x$. We choose one of the possible combinations (i.e. (i)) from both combinations. This led to the following 
\begin{align} 
&& x(t) \longrightarrow \; X ^{(b)} (t, \bar\vartheta)  = x(t) + \bar\vartheta\,(-\,g\,y\,\dot c)   
\; \equiv x(t) + \bar\vartheta\,[s_b\,x ],\nonumber\\
 && y (t) \longrightarrow \; Y ^{(b)} (t, \bar\vartheta)  = y (t) + \bar\vartheta\, (g\,x\,\dot c)
 \equiv  y(t) + \bar\vartheta\, [s_b\, y (t)]. 
\end{align} 
Similarly, we take other non-trivial BRST invariant restrictions $s_{b} (p_x\, c) = 0$, and  $s_{b} (p_y\, c) = 0$ and generalize  them onto a $(1, 1)$-dimensional supersub-manifold, we have:
\begin{align} 
 P_x (t, \bar\vartheta)\, F (t, \bar\vartheta)  = p_x (t) \, c(t) \Longrightarrow b_4 (t) \propto c (t) \quad \Longrightarrow \quad b_4 (t)  = \bar\kappa_1 \, c (t),  \nonumber\\
 P_y (t, \bar\vartheta)\, F (t, \bar\vartheta)  = p_y (t)\, c(t) \Longrightarrow b_5 (t) \propto c (t) \quad \Longrightarrow \quad b_5 (t)  = \bar\kappa_2 \, c (t). 
\end{align} 
 Again, to find the  values of $\bar\kappa_1$ and $\bar\kappa_2$, we use  the generalization of the BRST invariant restriction $s_b (p_x^2 + p_y^2) = 0$ as:
\begin{align} 
 P_x (t, \bar\vartheta)\, P_x (t, \bar\vartheta)  + P_y (t, \bar\vartheta)\, P_y (t, \bar\vartheta)  = x^2 (t) + y^2 (t) \; \Longrightarrow \; b_4 (t)\, P_x(t) + b_5 (t)\, p_y(t)  = 0.
\end{align} 
On substituting the values of $b_4$ and $b_5$ from Eq. (19), we obtain the relation: $\bar\kappa_1\,c\,x + \bar\kappa_2\,c\,y = 0$. This relation is valid for the two combinations of values: (i) $\bar\kappa_1 = -\,g\,y, \; \bar\kappa_2 = g\,p_x$ and   (ii) $\bar\kappa_1 = -\,g\,p_y, \; \bar\kappa_2 = g\,p_x$. We choose one of the possible combinations (i.e. (i)) from both combinations which leads to the following 
\begin{eqnarray} 
 &&p_x(t) \longrightarrow \; P_x ^{(b)} (t, \bar\vartheta)  = p_x(t) + \bar\vartheta\,(-\,g\,p_y\,\dot c)   
\; \equiv p_x(t) + \bar\vartheta\,[s_b\,p_x (t)],\nonumber\\
 &&p_y (t) \longrightarrow \; P_y ^{(b)} (t, \bar\vartheta)  = p_y (t) + \bar\vartheta\, (g\,p_x\,\dot c)
 \equiv  p_y(t) + \bar\vartheta\, [s_b\, p_y (t)]. 
\end{eqnarray} 
To find out the BRST transformations of variables $z$ and $c$, we use the generalization of the BRST invariant (0 + 1)-dimensional restrictions  [$s_b(z\,c) = 0$ and $s_b(\zeta\,\dot c) = 0$] onto (1, 1)-dimensional super sub-manifold as
\begin{align} 
Z (t, \bar\vartheta) \, F (t, \bar\vartheta)  = z (t)\, c(t) \Longrightarrow b_3 \propto c(t) \Longrightarrow  b_3  = m_1\, c(t),  \nonumber\\
\Xi (t, \bar\vartheta) \, \dot F (t, \bar\vartheta)  = \zeta (t)\, \dot c(t) \Longrightarrow b_7 \propto \dot c(t) \Longrightarrow  b_7  = m_2\, \dot c(t).
\end{align} 
Now using the generalization of the BRST invariant quantity $s_{b} (\dot\zeta  - z) = 0$ with the  substitutions of  the values of $b_3$ and $b_7$ into the anti-chiral super expansions leads to 
\begin{align} 
\dot\Xi (t, \bar\vartheta)  - {Z} (t, \bar\vartheta)  = \dot \zeta  (t) - z (t) \Longrightarrow m_1 = m_2.
\end{align} 
Finally, to ascertain the constants' values, we extend the BRST invariant restrictions $s_{b} ({b} \, z + i\, {\bar c}\, c) = 0$ and $s_{b} ({\dot b} \,\zeta + i\,\dot {\bar c}\, \dot c) = 0$ onto a $(1, 1)$-dimensional supersub-manifold.
In the equations:
\begin{eqnarray} 
 {{B}} ^{(b)} (t, \bar\vartheta)  \,{Z} (t, \bar\vartheta)  
+ i\,{\bar F} (t, \bar\vartheta) \,  F ^{(b)} (t, \bar\vartheta) 
& = & {b} (t) \, z (t) + i\, {\bar c} (t)\, c (t) \nonumber\\
& \Longrightarrow &  {b}_3 (t) = m_2 \, {b} (t), \nonumber\\ 
 \dot {B}^{(b)}(t, \bar\vartheta)  \,\Xi  (t, \bar\vartheta)  + i\,\dot {\bar F} (t, \bar\vartheta) \,  \dot F ^{(b)} (t, \bar\vartheta) 
& = &  \dot {b} (t) \, \zeta  (t) + i\,  \dot {\bar c} (t)\, \dot c (t) \nonumber\\
& \Longrightarrow & {b}_7 (t) = m_1 \, \dot {b} (t).
\end{eqnarray}
We observe that $m_1 = m_2 = 1$, as derived from Eq. (22) and Eq. (23). This allows us to determine the values of the derived variables as follows: $f_1 (t) = \dot c (t)$, $f_2 (t) = c (t)$, and $b_2 (t) =   {B} (t)$. Therefore, the expansions for the anti-chiral supervariables become:
\begin{eqnarray}
&&z (t) \;\;\longrightarrow \;\;{Z} ^{(b)} (t, \bar\vartheta)  = z (t) + \bar\vartheta\, [\dot c (t)] 
\equiv z (t) + \bar\vartheta\, [s_b\, z (t)], \nonumber\\
&& \zeta  (t) \;\; \longrightarrow \;\; \Xi  ^{(b)}(t, \bar\vartheta)  = \zeta  (t) + \bar\vartheta\,[c (t)]
 \equiv \varphi  (t) + \bar\vartheta\,[s_{b} \,\zeta  (t)].
\end{eqnarray}
Thus it is clear that in the expansion of an anti-chiral supervariable the coefficients of the Grasmmanian variable $\bar\vartheta$ are the BRST symmetry transformations that lead to the derivation of the BRST symmetries of variables present in the Lagrangian (8).  From the derivations of the above BRST symmetry transformations, we establish a connection between the BRST symmetry transformation $(s_b)$ and the partial derivative $(\partial_{\bar\vartheta})$ on the anti-chiral super sub-manifold. This connection is defined by the mapping: $s_b\longleftrightarrow \partial_{\bar\vartheta}$. In simpler terms, the BRST transformation of any generic variable $\psi(t)$ equals the translation of the corresponding generic anti-chiral supervariable $\Psi^{(b)}(t, \bar\vartheta)$ along the $\bar \vartheta$-direction. This can be expressed mathematically as: $s_{b}\, \psi(t) = \frac{\partial}{\partial {\bar \vartheta}} \Psi^{(b)}(t, \bar \vartheta) = \partial_{\bar \vartheta}\, \Psi^{(b)}(t, \bar \vartheta)$.

\vskip 0.2cm

We are now progressing towards deriving the quantum anti-BRST symmetry transformations using the chiral supervariable approach. To do this, we extend the (0 + 1)-dimensional variables to the (1, 1)-dimensional super sub-manifold of a suitably chosen (1, 2)-dimensional supermanifold. The chiral super expansions of the ordinary variables are given by:
\begin{eqnarray}
&&x (t) \quad \; \longrightarrow \quad X (t, \vartheta) \,\;= \; x (t) \; + \vartheta\,\bar b_1 (t),\nonumber\\
&&y (t) \quad \;\longrightarrow \quad Y (t, \vartheta) \;\;= \; y (t) \; + \vartheta\, \bar b_2 (t),\nonumber\\
&&z (t) \quad \;\longrightarrow \quad Z (t, \vartheta) \;\,\;= \; z (t) \; +  \vartheta\, \bar b_3 (t),\nonumber\\
&&p_x (t)\quad  \longrightarrow \quad P_x (t, \vartheta) \;= \; p_x (t) + \vartheta\,\bar b_4 (t),\nonumber\\
&&p_y (t) \quad \longrightarrow \quad P_y (t, \vartheta) \;= \; p_y (t) + \vartheta\, \bar  b_5 (t),\nonumber\\
&&p_z (t)  \quad\longrightarrow \quad P_z (t, \vartheta) \;= \; p_z (t)  + \vartheta\,\bar b_6 (t),\nonumber\\
&&\zeta (t) \,\;\quad \longrightarrow \quad \Xi (t, \vartheta) \;\,\;= \; \zeta (t) \; + \vartheta\,\bar b_7 (t),\nonumber\\
&&p_\zeta (t) \quad \longrightarrow \quad P_\zeta (t, \vartheta) \,\;= \; p_\zeta (t) + \bar\vartheta\,\bar b_8 (t),\nonumber\\
&&b (t) \quad \;\;\longrightarrow \quad B (t, \vartheta) \,\;= \; b (t) \;\;  + \vartheta\,\bar b_9 (t),\nonumber\\
&&c (t)  \;\;\quad\longrightarrow \quad F (t, \vartheta) \,\;= \; c (t) \;\;   + \vartheta\,\bar f_1 (t),\nonumber\\
&&\bar c (t) \;\quad \;\longrightarrow \quad  \bar F (t, \vartheta) \,\;= \; \bar c (t)\;\;   + \vartheta\,\bar f_2 (t),
\end{eqnarray}
where the secondary variables $\bar b_1, \bar b_2, \bar b_3, \bar b_4, \bar b_5, \bar b_6, \bar b_7, \bar b_8, \bar b_9$ are bosonic variables, and $\bar f_1, \bar f_2$ are the fermionic variables. The anti-BRST invariant restrictions must remain independent of the Grassmannian variable ($\vartheta$) when generalized onto the (1, 1)-dimensional chiral super sub-manifold. These restrictions are:
\begin{eqnarray}
&& s_{ab}(\bar c, b, p_z, p_\xi) = 0, \quad s_{ab} (x\, \bar c) = 0,\quad s_{ab} (y\, \bar c) = 0, \quad s_{ab} (x^2 + y^2) = 0, \quad s_{ab} (p_x\, \bar c) = 0, \quad  \nonumber\\
&&s_{ab} (p_y\, \bar c) = 0,\quad s_{ab} (p_x^2 + p_y^2) = 0,\quad s_{ab} (z\,\bar c) = 0, \quad s_{ab} (\zeta\, \dot {\bar c}) = 0,\nonumber\\
&& s_{ab} (\dot\zeta  - z) = 0, \quad s_{ab} (\dot b \, \zeta + i\, \dot {\bar c}\,\dot c) = 0,\quad s_{ab} (b\,z + i\,\bar c\, c) = 0.
\end{eqnarray}
To generalize these anti-BRST invariant restrictions onto the (1, 1)-dimensional super sub-manifold of the
most common (1, 2)-dimensional supermanifold, we express them as follows:
\begin{eqnarray} 
&& \bar F (t, \vartheta)  = \bar c (t), \; {B} (t, \vartheta)  = b (t), \;  P_z  (t, \vartheta)  = p_z  (t),\;
 P_\zeta  (t, \vartheta)  = p_\zeta  (t),\quad X (t, \vartheta)\, \bar F (t, \vartheta)  =\nonumber\\
 &&  x (t) \, \bar c(t), \quad 
 Y (t, \vartheta)\, \bar F (t, \vartheta)  = y (t) \, \bar c(t),\quad X^2 (t, \vartheta) + Y^2 (t, \vartheta)  = x^2 (t) + y^2 (t),\nonumber\\
 && P_x (t, \vartheta)\, \bar F (t, \vartheta) = p_x (t)\, \bar c (t), \quad P_y (t, \vartheta)\, F (t, \vartheta) = p_y (t)\,  c(t), \quad P_x^2 (t, \vartheta) + P_y^2 (t, \vartheta)  = \nonumber\\
 && p_x^2 (t) + p_y^2 (t), \quad {Z} (t, \vartheta) \, \bar F (t, \vartheta)  = z (t)\, \bar c(t), \quad \Xi (t, \vartheta) \, \dot {\bar F} (t, \vartheta)  
= \zeta (t)\, \dot {\bar c}(t),\nonumber\\
&&\dot\Xi (t, \vartheta)  - {Z} (t, \vartheta)  = \dot \zeta  (t) - z (t), \quad {\dot {{B}}} (t, \vartheta)  \,\Xi (t, \vartheta)  +  i\,\dot{\bar F} (t, \vartheta) \, \dot F (t, \vartheta) 
=  \nonumber\\
&&\dot  b (t) \, \xi (t) + i\, \dot {\bar c} (t)\,\dot {\bar c} (t), \quad { {B}} (t, \vartheta)  \,Z  (t, \vartheta)  + i\,{\bar F} (t, \vartheta) \,  F (t, \vartheta) 
=  {b} (t) \, z  (t) + i\,  {c} (t)\, \bar c (t). 
\end{eqnarray}
The above generalizations of the anti-BRST invariant restrictions [Eq. (16)] lead to the derivation of the chiral secondary variables, in a similar manner as done in the case of BRST symmetry transformations,  in terms of the auxiliary and basic variables present in the  Lagrangian $L$, which are as follows:
\begin{eqnarray}
 && \bar b_1 =  -\,g\,{y\,\bar c},\;\;  \bar b_2  = g\,x\,\bar c, \;\; b_3 = \bar c, \;\; \bar b_4 = -\,g\,p_y\,\bar c,\;\;\bar b_5  = g\,p_x\,\bar c,\nonumber\\
&& \bar b_6  = 0, \;\; \bar b_7  = \dot {\bar c},\;\; \bar b_8  = 0,\;\; \bar b_9 = 0,\;\; \bar f_1 
  = -\,i\,b,\;\; \bar f_2 = \bar c. 
\end{eqnarray}
The chiral secondary variables follow a derivation procedure similar to that of the anti-chiral derived variables.
Upon substituting these derived variables into the chiral super expansions (30), we obtain expressions
for the chiral supervariables on the (1, 1)-dimensional super sub-manifold:
\begin{eqnarray}
&&x (t) \quad \; \longrightarrow \quad X (t, \vartheta) \,\;= \; x (t) \; + \vartheta\,[-\,g\,{y\,\bar c}] 
\;\; \; \equiv \; x (t) \; + \vartheta\,[s_{ab} x(t)] ,\nonumber\\
&&y (t) \quad \;\longrightarrow \quad Y (t, \vartheta) \;\;= \; y (t) \; + \vartheta\, [g\,x\,\bar c]
\;\; \quad \; \equiv \; y (t) \; + \vartheta\, [s_{ab} y (t)],\nonumber\\
&&z (t) \quad \;\longrightarrow \quad Z (t, \vartheta) \;\,\;= \; z (t) \; +  \vartheta\, [\bar c]
\qquad \; \; \; \; \equiv \;  z (t) \; +  \vartheta\, [s_{ab} z (t)],\nonumber\\
&&p_x (t)\quad  \longrightarrow \quad P_x (t, \vartheta) \;= \; p_x (t) + \vartheta\,[-\,g\,p_y\,\bar c]
\; \equiv \;  p_x (t) + \vartheta\,[s_{ab} p_x (t)],\nonumber\\
&&p_y (t) \quad \longrightarrow \quad P_y (t, \vartheta) \;= \; p_y (t) + \vartheta\, [g\,p_x\,\bar c]
\quad\;  \equiv \; p_y (t) + \vartheta\, [s_{ab} p_y (t)],\nonumber\\
&&p_z (t)  \quad\longrightarrow \quad P_z (t, \vartheta) \;= \; p_z (t)  + \vartheta\,[0]
\qquad \;\; \; \; \equiv \;  p_z (t)  + \vartheta\,[s_{ab} p_z (t)],\nonumber\\
&&\zeta (t) \,\;\quad \longrightarrow \quad \Xi (t, \vartheta) \;\,\;= \; \zeta (t) \; + \vartheta\,[\dot {\bar c}]
\qquad\quad\; \equiv \; \zeta (t) \; + \vartheta\,[s_{ab} \zeta (t)],\nonumber\\
&&p_\zeta (t) \quad \longrightarrow \quad P_\zeta (t, \vartheta) \,\;= \; p_\zeta (t) + \vartheta\,[0]
\qquad\quad \equiv \; p_\zeta (t) + \vartheta\,[s_{ab} p_\zeta (t)],\nonumber\\
&&b (t) \quad \;\;\longrightarrow \quad B (t, \vartheta) \,\;= \; b (t) \;\;  + \vartheta\,[0]
 \qquad\quad\, \equiv \;  b (t) \;\;  + \vartheta\,[s_{ab} b(t)],\nonumber\\
&&c (t)  \;\;\quad\longrightarrow \quad F (t, \vartheta) \,\;= \; c (t) \;\;   + \vartheta\,[-\,i\,b] \;\;
\quad \; \equiv \;  c (t) \;\;   + \vartheta\,[s_{ab} c(t)],\nonumber\\
&&\bar c (t) \;\quad \;\longrightarrow \quad  \bar F (t, \vartheta) \,\;= \; \bar c (t)\;\;   + \vartheta\,[\bar c]
\qquad\quad\; \equiv \; \bar c (t)\;\;   + \vartheta\,[s_{ab} \bar c(t)].
\end{eqnarray}
In the above expressions, the coefficients of $\vartheta$ represent the anti-BRST symmetry transformations. Essentially, the anti-BRST symmetry transformation of any generic variable $\psi (t)$ corresponds to the translation of the corresponding generic 
chiral supervariable $\Psi^{(ab)}(t, \vartheta)$ along the $\vartheta$-direction [27-29]. This relationship is expressed 
mathematically as $s_{ab} \psi(t) = \frac{\partial}{\partial \vartheta} \Psi^{(ab)}(t, \vartheta)   
= \partial_\vartheta \Psi^{(ab)} (t, \vartheta)$. Thus, a mapping exists between the quantum anti-BRST 
symmetry transformation $(s_{ab})$ and the Grassmannian partial derivative $(\partial_{\vartheta})$ defined 
on the chiral supersub-manifold, denoted as $s_{ab} \longleftrightarrow  \partial_{\vartheta}$.


\section{Nilpotent (Anti-)co-BRST Symmetry Transformations: (Anti-)Chiral Supervariable Approach}

\vskip 0.2cm

In this section, we derive the nilpotent co-BRST and anti-co-BRST symmetry transformations (11) by exploiting the standard techniques of 
(anti-)chiral supervariable approach to BRST formalism.  We use chiral super expansions (30) and co-BRST invariant restrictions for the 
derivation of co-BRST symmetries, whereas, for the derivation of anti-co-BRST symmetries, we use anti-chiral super expansions and anti-BRST 
invariant restrictions. First of all, we derive the  co-BRST symmetry transformations using the following co-BRST invariant restrictions on the (1 + 1)-dimensional  super sub-manifold 
\begin{eqnarray}
&& s_d(\bar c, b, p_z, p_\zeta) = 0, \quad s_{d} (x\, \dot {\bar c}) = 0,\quad s_d (y\, \dot {\bar c}) = 0, \quad s_d (x^2 + y^2) = 0, \quad s_d (p_x\, \dot {\bar c}) = 0, \nonumber\\
&&s_d (p_y\, \dot {\bar c}) = 0,\quad s_d (p_x^2 + p_y^2) = 0,\quad s_d (z\,\dot {\bar c}) = 0, \quad s_d (\zeta\, \bar c) = 0,\quad  s_{d} (\dot\zeta  - z) = 0, \nonumber\\
&&s_{d} [\{g\,(x\,p_y - y\,p_x) + p_z\}\,z + i\, \dot {\bar c}\, c] = 0,\quad s_d [\{g\,(x\,p_y - y\,p_x) + p_z\}\,\zeta  + i\,\bar c\, c] = 0.
\end{eqnarray}
We generalize the above co-BRST invariant restrictions onto the (1, 1)-dimensional chiral super sub-manifolds (of the suitably chosen common (1, 2)-dimensional super-manifold)
\begin{eqnarray*} 
&& \bar F (t, \vartheta)  = \bar c (t), \; {B} (t, \vartheta)  = b (t), \;  P_z  (t, \vartheta)  = p_z  (t),\;
 P_\zeta  (t, \vartheta)  = p_\zeta  (t),\quad X (t, \vartheta)\, \bar F (t, \vartheta)  =\nonumber\\
 \end{eqnarray*}
 \begin{eqnarray} 
 &&  x (t) \, \bar c(t), \quad 
 Y (t, \vartheta)\, \bar F (t, \vartheta)  = y (t) \, \bar c(t),\quad X^2 (t, \vartheta) + Y^2 (t, \vartheta)  = x^2 (t) + y^2 (t),\nonumber\\
 && P_x (t, \vartheta)\, \bar F (t, \vartheta) = p_x (t)\, \bar c (t), \quad P_y (t, \vartheta)\, F (t, \vartheta) = p_y (t)\,  c(t), \quad P_x^2 (t, \vartheta) + P_y^2 (t, \vartheta)  = \nonumber\\
 && p_x^2 (t) + p_y^2 (t), \quad {Z} (t, \vartheta) \, \dot {\bar F} (t, \vartheta)  = z (t)\,\dot {\bar c}(t), \quad \Xi (t, \vartheta) \, {\bar F} (t, \vartheta)  
= \zeta (t)\, {\bar c}(t),\nonumber\\
&&\dot\Xi (t, \vartheta)  - {Z} (t, \vartheta)  = \dot \zeta  (t) - z (t), \quad [g\,(X (t, \vartheta)\,P_y (t, \vartheta) - Y (t, \vartheta)\,P_x (t, \vartheta)) +\nonumber\\
&&  P_z (t, \vartheta)]\,Z (t, \vartheta) + i\, \dot {\bar F} (t, \vartheta)\, C (t, \vartheta)  = [g\,(x (t)\,p_y (t) - y (t)\,p_x (t)) + p_z (t)]\,z (t) + i\, \dot {\bar c} (t)\, c (t),
 \nonumber\\
&& [g\,(X (t, \vartheta) \,P_y (t, \vartheta) - Y (t, \vartheta)\,P_x (t, \vartheta)) + P_z (t, \vartheta)]\,\Xi (t, \vartheta)  + i\,\bar F (t, \vartheta)\, F (t, \vartheta) \nonumber\\
&& =  [g\,(x (t)\,p_x (t) - y (t)\,p_x (t)) + p_z (t)]\,\zeta (t)  + i\,\bar c (t)\, c (t). 
\end{eqnarray}
At this point, we calculate the values of the derived variables in Equation (30) by applying the previously mentioned generalizations of the co-BRST invariant restrictions. To obtain these values, we start by considering the first line of Equation (36), where the generalized trivial co-BRST invariant quantities are introduced. This leads to the following relationships:
\begin{eqnarray} 
&&  \bar F (t, \vartheta)  = \bar c(t) \Longrightarrow f_2 = 0, 
\qquad\; \;\; \, P_z  (t, \vartheta)  = p_z  (t) \Longrightarrow b_6 = 0, \nonumber\\
&& P_\zeta  (t, \vartheta)  = p_\zeta  (t) \; \;\Longrightarrow b_8 = 0, \qquad {B} (t, \vartheta)  = b (t) \Longrightarrow b_9 = 0.
\end{eqnarray} 
By substituting the derived variable values into the expressions for the chiral super expansions in Equation (30), we arrive at the following chiral super expansions:
\begin{eqnarray}
&&\bar c (t) \; \, \longrightarrow \; \bar F ^{(d)} (t, \vartheta)  = \bar c(t) + \vartheta\,(0) \; \equiv c(t) + \vartheta\,[s_d\, c (t)] ,\nonumber\\
&& p_z (t) \longrightarrow \; P_z  ^{(d)} (t, \vartheta)  = p_z (t) + \vartheta\, (0) \equiv p_z (t)
 + \vartheta\, [s_d\, p_z (t)], \nonumber\\
&& p_\xi  (t)  \longrightarrow \; P_\xi  ^{(d)} (t, \vartheta)  =  p_\xi  (t) + \vartheta\, (0)
 \equiv p_\xi  (t) + \vartheta\, [s_d\, p_\xi  (t)], \nonumber\\ 
&&  b (t) \; \; \longrightarrow \; {{B}} ^{(d)} (t, \vartheta)  =   b(t) + \vartheta\,(0)\; \equiv  b (t) + \vartheta\,[s_d\, {b} (t)].
\end{eqnarray}
The superscript $(d)$ on the chiral supervariables denotes the supervariables obtained after the application of the co-BRST (i.e., dual-BRST) invariant restrictions. 
In the non-trivial case, we first generalize the co-BRST invariant restrictions $s_d(x\,\dot {\bar{c}}) = 0$ and $s_d(y\, \dot{\bar c}) = 0$ onto the (1,1)-dimensional 
super sub-manifold as follows:
\begin{eqnarray}
&& X(t, \vartheta) \,\dot {\bar F}(t, \vartheta) = x(t)\,\dot {\bar c} (t) \Longrightarrow b_1(t) \propto c(t) \Longrightarrow b_1(t) = \lambda_{1}\; \dot {\bar c} (t),
\nonumber\\
&& Y(t, \vartheta) \dot {\bar F} (t, \vartheta) = y(t) \,\dot {\bar c} (t) \; \Longrightarrow b_2(t) \propto c(t) \Longrightarrow b_2(t) = \lambda_{2}\; \dot {\bar c} (t).
\end{eqnarray}
To determine the values of $\lambda_1$ and $\lambda_2$, we generalize the co-BRST invariant restriction $s_b(x^2 + y^2) = 0$ as follows:
\begin{eqnarray}
X(t, \vartheta) \, X(t, \vartheta) + Y(t, \vartheta) \,Y(t, \vartheta) = x^2(t) + y^2(t) \Longrightarrow  b_1(t)\, x(t) + b_2(t)\, y(t) = 0.
\end{eqnarray}
Substituting the expressions for \(b_1\) and \(b_2\) into the equation, we get \( \lambda_1\, c\, x + \lambda_2\, c\, y = 0 \). 
This relation is satisfied for the two combinations of values: (i) \(\lambda_1 = -g\, y\), \(\lambda_2 = g\, x\) and (ii) \(\lambda_1 = g\, y\), \(\lambda_2 = -g\, x\). 
We choose the first combination (i), which results in:
\begin{eqnarray}
&& x(t) \longrightarrow  X^{(d)}(t, \vartheta) = x(t) + \vartheta \, [-g\, y\, \dot{c}] \equiv x(t) + \bar\vartheta \, [s_d x (t)],\nonumber\\
&& y(t) \longrightarrow  Y^{(d)}(t, \vartheta) = y(t) + \vartheta \,[g\, x\, \dot{c}] \equiv y(t) + \vartheta \,[s_d y(t)].
\end{eqnarray}
Similarly, by applying the co-BRST invariant conditions \( s_d(p_x\, \dot c) = 0 \) and \( s_d (p_y\, \dot c) = 0 \) and generalizing these onto a $((1,1)$-dimensional supersubmanifold, we obtain:
\begin{eqnarray}
&& P_x(t, \bar\vartheta) \dot F(t, \vartheta) = p_x(t) \dot c(t) \Longrightarrow b_4(t) \propto \dot c(t) \Longrightarrow  b_4(t) = \bar{\lambda}_1 \, \dot c(t),\nonumber\\
&& P_y(t, \bar\vartheta) \dot F(t, \vartheta) = p_y(t) \dot c(t) \Longrightarrow  b_5(t) \propto \dot c(t) \Longrightarrow b_5(t) = \bar{\lambda}_2 \, \dot c(t).
\end{eqnarray}
To find \(\bar{\lambda}_1\) and \(\bar{\lambda}_2\), we generalize \( s_d\,(p_x^2 + p_y^2) = 0 \) as follows:
\begin{eqnarray}
P_x(t, \vartheta) P_x(t, \vartheta) + P_y(t, \vartheta) P_y(t, \vartheta) = p_x^2(t) + p_y^2(t) \Longrightarrow  b_4(t)p_x(t) + b_5(t)p_y(t) = 0.
\end{eqnarray}
Substituting \( b_4 \) and \( b_5 \), we get \( \bar{\lambda}_1 \,\dot c \,p_x + \bar{\lambda}_2 \, \dot c \, p_y = 0 \), leading to the two possible combinations: (i) \(\bar{\lambda}_1 = -g\, p_y\), \(\bar{\lambda}_2 = g\, p_x\) and (ii) \(\bar{\lambda}_1 = g\, p_y\), \(\bar{\lambda}_2 = -g\, p_x\). We choose a combination (i), giving:
\begin{eqnarray}
&& p_x(t) \longrightarrow P_x^{(b)}(t, \vartheta) = p_x(t) + \vartheta (-g\, p_y\, \dot{c}) \equiv p_x(t) + \vartheta \, [s_d \, p_x],\nonumber\\
&& p_y(t) \longrightarrow P_y^{(b)}(t, \vartheta) = p_y(t) + \vartheta (g\, p_x\, \dot{c}) \equiv p_y(t) + \vartheta \, [s_d \, p_y].
\end{eqnarray}
Now to determine the co-BRST transformations of \( z \) and \( c \), we apply the co-BRST invariant restrictions \( s_d(z\,\dot {c}) = 0 \) and \( s_d(\zeta\,{c}) = 0 \) to the supersubmanifold:
\begin{eqnarray}
Z(t, \bar\vartheta) \dot {F}(t, \bar\vartheta) = z(t) \, \dot {c}(t) \Longrightarrow  b_3 \propto \dot c(t) \Longrightarrow b_3 = n_1 \, \dot c(t),\nonumber\\
\Xi(t, \bar\vartheta) {F}(t, \bar\vartheta) = \zeta(t) \,  {c}(t) \Longrightarrow b_7 \propto {c}(t) \Longrightarrow  b_7 = n_2 \,{c}(t).
\end{eqnarray} 
Using the co-BRST invariant condition \(s_d(\dot{\zeta} - z) = 0 \), and substituting \( b_3 \) and \( b_7 \), we obtain:
\begin{eqnarray}
\dot{\Xi}(t, \vartheta) - Z(t, \vartheta) = \dot{\zeta}(t) - z(t) \Longrightarrow  n_1 = n_2.
\end{eqnarray}
To fix the constant values $n_1$ and $n_2$, we again  generalize the co-BRST invariant constraints $s_{d} [\{g\,(x\,p_y - y\,p_x) + p_z\}\,z + i\, \dot {\bar c}\, c] = 0,\; s_d [\{g\,(x\,p_y - y\,p_x) + p_z\}\,\zeta  + i\,\bar c\, c] = 0,$ onto the super sub-manifold:
\begin{eqnarray}
&& [g\,(X (t, \vartheta)\,P_y (t, \vartheta) - Y (t, \vartheta)\,P_x (t, \vartheta)) +
P_z (t, \vartheta)]\,Z (t, \vartheta) + i\, \dot {\bar F} (t, \vartheta)\, c (t, \vartheta) \nonumber\\
&& = [g\,(x (t)\,p_y (t) - y (t)\,p_x (t)) + p_z (t)]\,z (t) + i\, \dot {\bar c} (t)\, c (t) \Longrightarrow \dot{b}_3(t) = n_2 \,\dot{b}(t),\nonumber\\
&& [g\,(X (t, \vartheta) \,P_y (t, \vartheta) - Y (t, \vartheta)\,P_x (t, \vartheta)) + P_z (t, \vartheta)] 
\,\Xi (t, \vartheta)  + i\,\bar F (t, \vartheta)\, F (t, \vartheta) \nonumber\\
&& =  [g\,(x (t)\,p_x (t) - y (t)\,p_x (t)) + p_z (t)]\,\zeta (t)  + i\,\bar c (t)\, c (t) \Longrightarrow b_7(t) = n_1 \,b(t).
\end{eqnarray}
From equations (45) and (47), we find \(n_1 = n_2 = 1\), allowing us to determine that \(f_1(t) = \dot{c}(t)\), \(f_2(t) = c (t)\), and \(b_2(t) = \mathcal{B}(t)\). Hence, the expansions of the anti-chiral supervariables become:
\begin{eqnarray}
z(t) \longrightarrow Z^{(d)}(t, \vartheta) = z(t) + \vartheta \,[\dot{c}(t)] \equiv z(t) + \vartheta [s_b z(t)],\nonumber\\
\zeta(t) \longrightarrow \Xi^{(d)}(t, \vartheta) = \zeta(t) + \vartheta \,[c(t)] \equiv \zeta(t) + \vartheta [s_b \zeta(t)].
\end{eqnarray}
The expansion of a chiral supervariable reveals that the coefficients of the Grassmann variable $\vartheta$ correspond to the co-BRST symmetry transformations. This leads to the derivation of the co-BRST symmetries for the variables in the Lagrangian (8). From this derivation, a connection between the co-BRST symmetry transformation $(s_d)$ and the partial derivative $(\partial_{\vartheta})$ on the chiral super sub-manifold is established. This relationship is expressed by the mapping: $s_d \longleftrightarrow \partial_{\vartheta}$. In simpler terms, the co-BRST transformation of any generic variable $\psi(t)$ corresponds to the translation of the related chiral supervariable $\Psi^{(d)}(t, \vartheta)$ in the $\vartheta$-direction. Mathematically, this is expressed as: 
\begin{eqnarray}
s_{d} \, \psi(t) = \frac{\partial}{\partial {\vartheta}} \Psi^{(d)}(t, {\vartheta}) = \partial_{ \vartheta} \Psi^{(d)}(t, \vartheta).
\end{eqnarray}
We are now advancing towards deriving the quantum anti-co-BRST symmetry transformations through the anti-chiral supervariable technique. In this framework, we extend the (0 + 1)-dimensional variables to a (1, 1)-dimensional super sub-manifold embedded within an appropriately selected (1, 2)-dimensional supermanifold. The anti-chiral supervariable expansions of the ordinary variables take the form:
\begin{eqnarray}
&&x (t) \quad \; \longrightarrow \quad X (t, \bar\vartheta) \,\;= \; x (t) \; + \bar\vartheta\, b_1 (t),\nonumber\\
&&y (t) \quad \;\longrightarrow \quad Y (t, \bar\vartheta) \;\;= \; y (t) \; + \bar\vartheta\,  b_2 (t),\nonumber\\
&&z (t) \quad \;\longrightarrow \quad Z (t, \bar\vartheta) \;\,\;= \; z (t) \; +  \bar\vartheta\,  b_3 (t),\nonumber\\
&&p_x (t)\quad  \longrightarrow \quad P_x (t, \bar\vartheta) \;= \; p_x (t) + \bar\vartheta\, b_4 (t),\nonumber\\
&&p_y (t) \quad \longrightarrow \quad P_y (t, \bar\vartheta) \;= \; p_y (t) + \bar\vartheta\,  b_5 (t),\nonumber\\
&&p_z (t)  \quad\longrightarrow \quad P_z (t, \bar\vartheta) \;= \; p_z (t)  + \bar\vartheta\, b_6 (t),\nonumber\\
&&\zeta (t) \,\;\quad \longrightarrow \quad \Xi (t, \bar\vartheta) \;\,\;= \; \zeta (t) \; + \bar\vartheta\, b_7 (t),\nonumber\\
&&p_\zeta (t) \quad \longrightarrow \quad P_\zeta (t, \bar\vartheta) \,\;= \; p_\zeta (t) + \bar\vartheta\, b_8 (t),\nonumber\\
&&b (t) \quad \;\;\longrightarrow \quad B (t, \bar\vartheta) \,\;= \; b (t) \;\;  + \bar\vartheta\, b_9 (t),\nonumber\\
&&c (t)  \;\;\quad\longrightarrow \quad F (t, \bar\vartheta) \,\;= \; c (t) \;\;   + \vartheta\,f_1 (t),\nonumber\\
&&\bar c (t) \;\quad \;\longrightarrow \quad  \bar F (t, \bar\vartheta) \,\;= \; \bar c (t)\;\;   + \bar\vartheta\, f_2 (t).
\end{eqnarray}
Here $b_1, b_2, b_3,  b_4,  b_5,  b_6,  b_7, b_8,  b_9$ are the {fermionic}  secondary variables, while $f_1$ and $f_2$ are the {bosonic}  
secondary variables. To ensure that the anti-BRST invariant restrictions remain unaffected by the Grassmannian coordinate ($\bar\vartheta$) 
when generalized to the (1, 1)-dimensional chiral super sub-manifold, we impose the following restrictions:
\begin{eqnarray}
&& s_{ad}(c, b, p_z, p_\zeta) = 0, \quad s_{ad} (x\, \dot {c}) = 0,\quad s_{ad} (y\, \dot {c}) = 0, \quad s_{ad} (x^2 + y^2) = 0, \quad s_{ad} (p_x\, \dot {c}) = 0, \nonumber\\
&&s_{ad} (p_y\, \dot {c}) = 0,\quad s_{ad} (p_x^2 + p_y^2) = 0,\quad s_{ad} (z\,\dot {c}) = 0, \quad s_{ad} (\zeta\, c) = 0,\quad  s_{ad} (\dot\zeta  - z) = 0, \nonumber\\
&&s_{ad} [\{g\,(x\,p_x - y\,p_x) + p_z\}\,z - i\, {\bar c}\, \dot c] = 0,\quad s_{ad} [\{g\,(x\,p_x - y\,p_x) + p_z\}\,\zeta   i\,\bar c\, c] = 0.
\end{eqnarray}
To extend these anti-co-BRST invariant restrictions onto the (1, 1)-dimensional super sub-manifold of the most general 
(1, 2)-dimensional supermanifold, we express these crucial  anti-BRST restrictions  in the following form: 
\begin{eqnarray*}
&& F (t, \bar\vartheta)  = c (t), \; {B} (t, \bar\vartheta)  = b (t), \;  P_z  (t, \bar\vartheta)  = p_z  (t),\;
 P_\zeta  (t, \bar\vartheta)  = p_\zeta  (t),\quad X (t, \bar\vartheta)\, \dot F (t, \bar\vartheta)  =\nonumber\\
 &&  x (t) \, \dot c(t), \quad 
 Y (t, \bar\vartheta)\, \dot F (t, \bar\vartheta)  = y (t) \, \dot c(t),\quad X^2 (t, \bar\vartheta) + Y^2 (t, \bar\vartheta)  = x^2 (t) + y^2 (t),\nonumber\\
 && P_x (t, \bar\vartheta)\, \dot F (t, \bar\vartheta) = p_x (t)\, \dot c (t), \quad P_y (t, \bar\vartheta)\, \dot F (t, \bar\vartheta) = p_y (t)\,  \dot c(t), \quad  \nonumber\\
 \end{eqnarray*}
 \begin{eqnarray} 
 && P_x^2 (t, \bar\vartheta) + P_y^2 (t, \bar\vartheta)  = p_x^2 (t) + p_y^2 (t), \quad {Z} (t, \bar\vartheta) \, \dot {F} (t, \bar\vartheta)  = z (t)\,\dot {c}(t),\nonumber\\
 && \Xi (t, \bar\vartheta) \, {F} (t, \bar\vartheta)  
= \zeta (t)\, {c}(t),\quad \dot \Xi (t, \bar\vartheta)  - {Z} (t, \bar\vartheta)  = \dot \zeta  (t) - z (t), \nonumber\\
&& [g\,(X (t, \bar\vartheta)\,P_x (t, \bar\vartheta) - Y (t, \vartheta)\,P_x (t, \bar\vartheta)) +  P_z (t, \bar\vartheta)]\,Z (t, \bar\vartheta) - i\, {\bar F} (t, \bar\vartheta)\, \dot  F (t, \bar\vartheta)\nonumber\\
&& = [g\,(x (t)\,p_x (t) - y (t)\,p_x (t)) + p_z (t)]\,z (t) - i\, {\bar c} (t)\, \dot c (t),
 \nonumber\\
&& [g\,(X (t, \bar\vartheta) \,P_x (t, \bar\vartheta) - Y (t, \bar\vartheta)\,P_x (t, \bar\vartheta)) + P_z (t, \bar\vartheta)]\,\Xi (t, \bar\vartheta)  - i\,\bar F (t, \bar\vartheta)\, F (t, \bar\vartheta) \nonumber\\
&& =  [g\,(x (t)\,p_x (t) - y (t)\,p_x (t)) + p_z (t)]\,\zeta (t)  - i\,\bar c (t)\, c (t). 
\end{eqnarray}
The above generalization of the anti-co-BRST invariant restrictions leads to the derivation of the chiral secondary variables,
similar to the procedure followed for co-BRST symmetry transformations, expressed in terms of auxiliary and basic variables in the Lagrangian $L$:
\begin{eqnarray}
 &&  b_1 =  -\,g\,y\,\dot c,\;\;  b_2  = g\,x\, \dot { c}, \;\; b_3 = \dot c, \;\;  b_4 = -\,g\,p_y\,\dot c,\;\;b_5  = g\,p_x\,\dot c,\nonumber\\
&& b_6  = 0, \;\; b_7  = \dot {c},\;\; b_8  = 0,\;\;  b_9 = 0,\;\;  f_1 
  = 0\;\;  f_2 =  i\,[g\,(x\,p_y - y\,p_x) + p_z].
\end{eqnarray}
The determination of secondary anti-chiral variables follows a similar process to that of the chiral secondary variables. 
Upon substituting these into the anti-chiral super expansions, we arrive at the expressions for the anti-chiral supervariables 
on the (1, 1)-dimensional super sub-manifold of common (1, 2)-dimensional supermanifold:
\begin{eqnarray}
&& x (t) \quad \; \longrightarrow \quad X (t, \bar\vartheta) \,\;= \; x (t) \; + \bar\vartheta\,[-\,g\,y\,\dot c] 
\;\; \; \equiv \;\; x (t) \; + \bar\vartheta\,[s_{ad}\, x(t)] ,\nonumber\\
&& y (t) \quad \;\longrightarrow \quad Y (t, \bar\vartheta) \;\;= \; y (t) \; + \bar\vartheta\, [g\,x\,  {\bar c}]
\;\; \quad \; \equiv\; \; y (t) \; + \bar\vartheta\, [s_{ad} \, y (t)],\nonumber\\
&& z (t) \quad \;\longrightarrow \quad Z (t, \bar\vartheta) \;\,\;= \; z (t) \; +  \bar\vartheta\, [\dot c]
\qquad \; \; \; \; \equiv \;\;  z (t) \; +  \bar\vartheta\, [s_{ad}\, z (t)],\nonumber\\
&& p_x (t)\quad  \longrightarrow \quad P_x (t, \bar\vartheta) \;= \; p_x (t) + \bar\vartheta\,[-\,g\,p_y\,\dot c]
\; \equiv \;  p_x (t) + \bar\vartheta\,[s_{ad}\, p_x (t)],\nonumber\\
&& p_y (t) \quad \longrightarrow \quad P_y (t, \bar\vartheta) \;= \; p_y (t) + \bar\vartheta\, [g\,p_x\,\dot c]
\quad\;  \equiv\; \; p_y (t) + \bar\vartheta\, [s_{ad}\, p_y (t)],\nonumber\\
&& p_z (t)  \quad\longrightarrow \quad P_z (t, \bar\vartheta) \;= \; p_z (t)  + \bar\vartheta\,[0]
\qquad \;\; \; \; \equiv \;\;  p_z (t)  + \bar\vartheta\,[s_{ad}\, p_z (t)],\nonumber\\
&& \zeta (t) \,\;\quad \longrightarrow \quad \Xi (t, \bar\vartheta) \;\,\;= \; \zeta (t) \; + \bar\vartheta\,[c]
\qquad\quad\; \equiv \;\; \zeta (t) \; + \bar\vartheta\,[s_{ad} \, \zeta (t)],\nonumber\\
&& p_\zeta (t) \quad \longrightarrow \quad P_\zeta (t, \vartheta) \,\;= \; p_\zeta (t) + \bar\vartheta\,[0]
\qquad\quad \equiv \;\; p_\zeta (t) + \bar\vartheta\,[s_{ad} \, p_\zeta (t)],\nonumber\\
&& b (t) \quad \;\;\longrightarrow \quad B (t, \vartheta) \,\;= \; b (t) \;\;  + \bar\vartheta\,[0]
 \qquad\quad\, \equiv \; \; b (t) \;\;  + \bar\vartheta\,[s_{ad}\, b(t)],\nonumber\\
&& c (t)  \;\;\quad\longrightarrow \quad F (t, \bar\vartheta) \,\;= \; c (t) \;\;   + \bar\vartheta\,[0]
\quad \qquad\, \equiv \;\;  c (t) \;\;   + \bar\vartheta\,[s_{ad}\, c(t)],\nonumber\\
&& \bar c (t) \;\quad \;\;\longrightarrow \quad  \bar F (t, \bar\vartheta) \,\;= \; \bar c (t)\;\;   + \bar\vartheta\,[i\,\{g\,(x\,p_y - y\,p_x) + p_z\}]\nonumber\\
 &&~~~~~~~~~~~~~~~~~~~~~~~~~~~~~~~~~~~~~~~~~~~~~~~~~~~~~~~~~~~ \,\equiv \;\; \bar c (t)\;\;   + \bar\vartheta\,[s_{ad}\, \bar c(t)].
\end{eqnarray}
In these expansions, the coefficients of $\bar\vartheta$ represent the anti-co-BRST symmetry transformations. In essence, 
the anti-co-BRST transformation of any generic variable $\psi (t)$ is associated with the translation of the corresponding 
chiral supervariable $\Psi^{(ad)}(t, \bar\vartheta)$ along the $\bar\vartheta$-direction
[27-29]. This relationship is mathematically expressed as $s_{ad}\, \psi(t) = \frac{\partial}{\partial_{\bar\vartheta}} \,\Psi^{(ad)}(t, \bar\vartheta) = 
\partial_{\bar\vartheta} \,\Psi^{(ad)} (t, \bar\vartheta)$. Hence, there is a mapping between the quantum anti-co-BRST symmetry transformation $(s_{ad})$ and the Grassmannian partial derivative $(\partial_{\bar\vartheta})$ defined on the chiral  supersubmanifold, denoted as $s_{ad} \longleftrightarrow  \partial_{\bar\vartheta}$.

\section{Nilpotency and Absolute Anti-Commutativity Properties of the Noether Conserved Charges: ACSA}

\vskip 0.2cm

In this section, we explore the nilpotency and absolute anti-commutativity of the conserved (anti-)BRST and (anti-)co-BRST charges
using the (anti-)chiral superfield Approach (ACSA). We start by demonstrating the nilpotency of these charges. It is straightforward
to represent the (anti-)BRST and (anti-)co-BRST charges in terms of (anti-)chiral supervariables and the partial derivatives 
$(\partial_{\bar\vartheta}, \partial_\vartheta)$, with the integral forms given as follows:
\begin{eqnarray}
{\cal Q}_{b} & = &  -\,i\,\frac {\partial}{\partial \bar \vartheta} \Big[{\bar F} ^ {(b)} (t, \bar \vartheta)\, \dot F ^ {(b)} (t, \bar \vartheta) 
-  \dot  {\bar F} ^ {(b)} (t, \bar \vartheta) \,  F ^ {(b)} (t, \bar \vartheta) \Big] \nonumber\\
& \equiv &  -\,i\,\int \,d \bar\vartheta \Big[{\bar F} ^ {(b)} (t, \bar \vartheta)\, \dot F ^ {(b)} (t, \bar \vartheta) 
-  \dot  {\bar F} ^ {(b)} (t, \bar \vartheta) \,  F ^ {(b)} (t, \bar \vartheta)\Big],\nonumber\\
{\cal Q}_{ab} & = & \frac {\partial}{\partial \vartheta}\, \Big[i\,{\bar F} ^ {(ab)} (t, \vartheta)\, \dot {F} ^ {(ab)} (t, \vartheta) 
-  i\,\dot  {\bar F} ^ {(ab)} (t, \vartheta) \, {\bar  F} ^ {(ab)} (t,  \vartheta)  \Big] \nonumber\\
&\equiv & \int \, d \vartheta\, \Big[i\,{\bar F} ^ {(ab)} (t, \vartheta)\, \dot {F} ^ {(ab)} (t, \vartheta) 
-  i\,\dot  {\bar F} ^ {(ab)} (t, \vartheta) \, {\bar  F} ^ {(ab)} (t, \vartheta)  \Big], 
\end{eqnarray}
\begin{eqnarray}
{\cal Q}_d &=& \frac {\partial}{\partial \vartheta}\, \Big[i\,{\bar F} ^ {(d)} (t, \vartheta)\, \dot {F} ^ {(d)} (t,  \vartheta) 
-  i\,\dot  {\bar F} ^ {(d)} (t, \bar \vartheta) \,  {\bar  F} ^ {(d)} (t, \vartheta)   \Big]\nonumber\\ 
&\equiv & \int \, d \vartheta \, \Big[i\,\bar F ^ {(d)} (t, \vartheta)   \, \dot F  ^ {(d)} (t, \vartheta)   
- i\, \dot {\bar F} ^ {(d)} (t, \vartheta)  \, F ^ {(d)}  (t, \vartheta)   \Big], \nonumber\\
{\cal Q}_{ad} &=& \frac {\partial}{\partial \bar \vartheta}\, \Big[i\, \dot {\bar F} ^ {(ad)}(t, \bar\vartheta) \, F ^ {(ad)}(t, \bar\vartheta)  
- i\,\bar F ^ {(ad)} (t, \bar\vartheta)  \, \dot F  ^ {(ad)}(t, \bar\vartheta)   \Big]\nonumber\\ 
&\equiv & \int \, d \bar\vartheta \, \Big[i\, \dot {\bar F} ^ {(ad)}(t, \bar\vartheta) \, F  ^ {(ad)}(t, \bar\vartheta)  
- i\,\bar F  ^ {(ad)}(t, \bar\vartheta)  \, \dot F  ^ {(ad)}(t, \bar\vartheta)   \Big].
\end{eqnarray}
The superscripts $(b)$ and $(ab)$ denote the {anti-chiral} and {chiral} supervariables, respectively, 
which are obtained by applying BRST and anti-BRST invariant restrictions. Similarly, the superscripts $(d)$ and $(ad)$
represent the {chiral} and {anti-chiral} supervariables, respectively, resulting from the application of
co-BRST and anti-co-BRST invariant restrictions. It is evident that the nilpotency conditions $(\partial_{\bar\vartheta}^2 = 0,
\; \partial_{\vartheta}^2 = 0)$ of the translational generators $(\partial_{\bar\vartheta}, \; \partial_{\vartheta})$ imply that
\begin{eqnarray}
&&\partial_{\bar\vartheta}\; {\cal Q}_{b}  = 0 
\quad\;\,\Longleftrightarrow \quad s_b\; {\cal Q}_{b}  \quad \; = \; -\;i\;{\{{\cal Q}_{b},   {\cal Q}_b}\} = 0,\nonumber\\
&&\partial_{\vartheta}\; {\cal Q}_{ab}  = 0
\quad\Longleftrightarrow \quad s_{ab}\; {\cal Q}_{ab}  \; \,\,= \; -\;i\;{\{{\cal Q}_{ab}, {\cal Q}_{ab}}\} = 0,\nonumber\\
&&\partial_{\vartheta}\; {\cal Q}_d  = 0 
\quad\;\Longleftrightarrow \quad s_d\; {\cal Q}_d  \; \quad = \; -\;i\;{\{{\cal Q}_d, {\cal Q}_d}\} = 0,\nonumber\\
&&\partial_{\bar\vartheta}\; {\cal Q}_{ad}  = 0
\quad\Longleftrightarrow \quad s_{ad}\; {\cal Q}_{ad}  \;\; = \; -\;i\;{\{{\cal Q}_{ad}, {\cal Q}_{ad}}\} = 0.
\end{eqnarray}
This demonstrates the nilpotency [$ {\cal Q}_{(a)b}^2 = {\cal Q}_{(a)d}^2 = 0 $] of the conserved charges within the framework of ACSA to BRST formalism. 
Hence, we have established a profound connection between the nilpotency $(\partial_{\bar\vartheta}^2 = 0, \;\partial_{\vartheta}^2 = 0)$ of the translational
generators $(\partial_{\bar\vartheta}, \;\partial_{\vartheta})$ and the nilpotency [i.e., $ {\cal Q}_{(a)b}^2 = {\cal Q}_{(a)d}^2 = 0 $] of the (anti-)BRST
and (anti-)co-BRST charges [$ {\cal Q}_{(a)b}, \; {\cal Q}_{(a)d} $]. This property of nilpotency can also be expressed in ordinary space using the (anti-)BRST
{exact} and (anti-)co-BRST {exact} forms of the charges, such as:
\begin{eqnarray}
{\cal Q}_{b} &=& -\, i\,s_b\, \big(\bar c \, \dot c - \dot {\bar c}\, c\big),  \qquad {\cal Q}_{ab}  
=  +\,i \,s_{ab}\, \big(\bar c \, \dot c - \dot {\bar c}\, c\big),\nonumber\\
{\cal Q}_d &=& +\,i\,s_d\, \big(\bar c \, \dot c - \dot {\bar c}\, c\big),  \qquad {\cal Q}_{ad} 
= - \,i \,s_{ad}\, \big(\bar c \, \dot c - \dot {\bar c}\, c\big). 
\end{eqnarray}
The above expressions show the nilpotency property of the (anti-)BRST along with (anti-) co-BRST conserved charges, in a simpler way, in an ordinary space [cf. (44)].

\vskip 0.2cm

We are now in a position to demonstrate the absolute anti-commutativity of the (anti-) BRST and (anti-)co-BRST charges. To achieve this, we express the charges in terms of the (anti-)chiral supervariables and the derivatives $(\partial_\vartheta, \; \partial_{\bar\vartheta})$ associated with the Grassmannian variables $(\bar\vartheta, \vartheta)$.
\begin{eqnarray}
{\cal Q}_{b} &=&  - \,i\,\frac {\partial}{\partial \vartheta}  \,\Big[\dot F ^ {(ab)} (t, \vartheta)    \, F ^ {(ab)} (t, \vartheta)  \Big]   
   \, \equiv   - \,i\,\int \, d \vartheta  \,\Big[\dot F ^ {(ab)} (t, \vartheta)    \, F ^ {(ab)} (t, \vartheta)  \Big],\nonumber\\
{\cal Q}_{ab} &=&  \;\;\; i\,\frac {\partial}{\partial \bar \vartheta}  \,\Big[\dot {\bar F} ^ {(b)} (t, \bar \vartheta)  \, \bar F ^ {(b)} (t, \bar \vartheta)\Big] \quad  \,\equiv  \;\; \; i\,\int \, d \bar \vartheta  \,\Big[\dot {\bar F } ^ {(b)} (t, \bar\vartheta)   \, \bar F ^ {(b)} (t, \bar\vartheta) \Big],\nonumber\\
{\cal Q}_d &=&  \;\; \; i\, \frac {\partial}{\partial \bar \vartheta} \,\Big[\dot {\bar F} ^ {(ad)} (t, \bar\vartheta)  \,
 \bar F ^ {(ad)} (t, \bar\vartheta) \Big] \, \,\equiv   \; \;\; i\, \int d \bar\vartheta \,\Big[\dot {\bar F} ^ {(ad)} (t, \bar\vartheta) 
 \, \bar F ^ {(ad)}(t, \bar\vartheta) \Big],\nonumber\\
{\cal Q}_{ad} &=&  -\,i\, \frac {\partial}{\partial  \vartheta} \,\Big[\dot {F} ^ {(d)} (t, \vartheta)   \, F ^ {(d)} (t, \vartheta)  \Big] 
 \quad \equiv   -\,i\, \int d \vartheta \,\Big[\dot {F} ^ {(d)} (t, \vartheta)   \, F ^ {(d)} (t, \vartheta)  \Big], 
\end{eqnarray}
where the superscripts $(a)b$ and $(a)d$ carry the same meaning as explained earlier. Here, it is straightforward to verify that the nilpotency $(\partial_{\bar\vartheta}^2 = 0, \;\partial_{\vartheta}^2 = 0)$ of the translational generators $(\partial_{\bar\vartheta}, \; \partial_{\vartheta})$ leads to the following relations. The superscripts $(a)b$ and $(a)d$ carry the same meaning as explained earlier. Here, it is straightforward to verify that the nilpotency $(\partial_{\bar\vartheta}^2 = 0, \;\partial_{\vartheta}^2 = 0)$ of the translational generators $(\partial_{\bar\vartheta}, \; \partial_{\vartheta})$ leads to the following relations:
\begin{eqnarray}
&&\partial_{\vartheta}\; {\cal Q}_{b}  = 0 \; \quad\, 
\Longleftrightarrow \quad s_{ab}\; {\cal Q}_{b}  = -\,i\;{\{{\cal Q}_{b},   {\cal Q}_{ab}}\} = 0,\nonumber\\
&&\partial_{\bar\vartheta}\;{\cal Q}_{ab} = 0 \quad 
\Longleftrightarrow \quad s_b\;{\cal Q}_{ab} = - \,i\;{\{{\cal Q}_{ab}, {\cal Q}_b}\} = 0,\nonumber\\
&&\partial_{\bar\vartheta}\; {\cal Q}_d  = 0 \; \quad 
\Longleftrightarrow \quad s_{ad}\; {\cal Q}_d  = -\,i\;{\{{\cal Q}_d, {\cal Q}_{ad}}\} = 0,\nonumber\\
&&\partial_{\vartheta}\;{\cal Q}_{ad} = 0 \quad 
\Longleftrightarrow \quad s_d\;{\cal Q}_{ad} = - \,i\;{\{{\cal Q}_{ad}, {\cal Q}_d}\} = 0,
\end{eqnarray}
which demonstrates the absolute anti-commutativity property of the (anti-)BRST as well as (anti-)co-BRST conserved 
charges. This property of absolute anti-commutativity of the conserved charges can also be explicitly shown 
in ordinary space by using the following (anti-)BRST {exact} and (anti-)co-BRST {exact} forms 
of the charges, namely:
\begin{eqnarray}
&& {\cal Q}_{b} = - \,i\, s_{ab} \,\big(\dot c \, c\big), \qquad {\cal Q}_{ab} =  +\,i\, s_{b} \,\big(\dot {\bar c} \, 
\bar c\big),  \nonumber\\
&& {\cal Q}_d  = +\,i\, s_{ad} \,\big(\dot {\bar{ c}} \, \bar c\big),  \qquad  {\cal Q}_{ad} =  
- \, i\, s_d \,\big(\dot c \,  c\big), 
\end{eqnarray}
which are the short forms of Noether's conserved charges [Eqs. (58), (61)] in terms of (anti-) BRST and (anti-)co-BRST symmetry transformations of ghost and anti-ghost fields. We end this section with the following concluding remarks. We show the nilpotency and absolute anti-commutativity properties of Noether's conserved charges  (10) and (13) by using only one Grassmannian variable out of two usual Grassmannian variables $(\vartheta, \, \bar\vartheta)$.

\section {Invariance of Lagrangian: ACSA}

\vskip 0.2cm

In this section, we discuss the (anti-)BRST and (anti-)co-BRST symmetries invariance of the Lagrangian within the framework of the (anti-)chiral supervariable approach (ACSA) to BRST formalism. To achieve this, we first extend the standard Lagrangian in (0 + 1)-dimensions to a suitably chosen (1, 1)-dimensional (anti-)chiral super sub-manifold of the familiar (1, 2)-dimensional supermanifold. The resulting expressions for the (anti-)chiral super Lagrangian are given as 
\begin{eqnarray*}
L(t) & \longrightarrow & \tilde L ^{(ac)} (t, \bar\vartheta) \nonumber\\
& = & P_x ^{(b)} (t, \bar\vartheta)\, \dot X ^{(b)} (t, \bar\vartheta) +  P_y ^{(b)} (t, \bar\vartheta)\, \dot Y  ^{(b)} (t, \bar\vartheta) +  P_z ^{(b)}  (t, \bar\vartheta)\, \dot Z ^{(b)} (t, \bar\vartheta)\nonumber\\
& + & \frac{1}{2}\, \big[P_x ^{(b)} (t, \bar\vartheta)\;P_x ^{(b)} (t, \bar\vartheta) + P_y ^{(b)} (t, \bar\vartheta)\;P_y ^{(b)} (t, \bar\vartheta) + P_z ^{(b)} (t, \bar\vartheta)\;P_z ^{(b)} (t, \bar\vartheta)\big]\nonumber\\
& - & \Xi ^{(b)} (t, \bar\vartheta)\; \big[g \,\{ X^{(b)} (t, \bar\vartheta)\; P_y ^{(b)} (t, \bar\vartheta) - Y ^{(b)} (t, \bar\vartheta)\; P_x ^{(b)} (t, \bar\vartheta)\}  +  P_z ^{(b)} (t, \bar\vartheta)\big]\nonumber\\
& + & B^{(b)} (t, \bar\vartheta) \; \big[\dot \Xi ^{(b)} (t, \bar\vartheta) - Z ^{(b)} (t, \bar\vartheta)\big] 
 + \frac{1}{2}\,B^{(b)} (t, \bar\vartheta)\,B^{(b)} (t, \bar\vartheta)  \nonumber\\
 & - & i\, \dot {\bar F } ^{(b)} (t, \bar\vartheta)\, F^{(b)} (t, \bar\vartheta)  -  i\, {\bar F}^{(b)} (t, \bar\vartheta)\, F^{(b)} (t, \bar\vartheta),
\end{eqnarray*}
\begin{eqnarray}
L(t) & \longrightarrow & \tilde L ^{(c)} (t, \vartheta)\nonumber\\
& = & P_x ^{(ab)} (t, \vartheta)\, \dot X ^{(ab)} (t, \vartheta) +  P_y ^{(ab)} (t, \vartheta)\, \dot Y  ^{(ab)} (t, \vartheta) +  P_z ^{(ab)}  (t, \vartheta)\, \dot Z ^{(ab)} (t, \vartheta)\nonumber\\
& + & \frac{1}{2}\, \big[P_x ^{(ab)} (t, \vartheta)\;P_x ^{(ab)} (t, \vartheta) + P_y ^{(ab)} (t, \vartheta)\;P_y ^{(ab)} (t, \vartheta) + P_z ^{(ab)} (t, \vartheta)\;P_z ^{(ab)} (t, \vartheta)\big]\nonumber\\
& - & \Xi ^{(ab)} (t, \vartheta)\; \big[g\, \{ X^{(ab)} (t, \vartheta)\; P_y ^{(ab)} (t, \vartheta) - Y ^{(ab)} (t, \vartheta)\; P_x ^{(ab)} (t, \vartheta)\}  +  P_z ^{(ab)} (t, \vartheta)\big]\nonumber\\
& + & B^{(ab)} (t, \vartheta) \; \big[\dot \Xi ^{(ab)} (t, \vartheta) - Z ^{(ab)} (t, \vartheta)\big] 
 + \frac{1}{2}\,B^{(ab)} (t, \vartheta)\,B^{(ab)} (t, \vartheta)  \nonumber\\
 & - & i\, \dot {\bar F } ^{(ab)} (t, \vartheta)\, F^{(ab)} (t, \vartheta)  -  i\, {\bar F}^{(ab)} (t, \vartheta)\, F^{(ab)} (t, \vartheta).
\end{eqnarray}
Here, the superscripts 
$(ac)$ and $(c)$ on the super Lagrangians refer to the anti-chiral and chiral super Lagrangians, respectively, 
which incorporate the corresponding anti-chiral and chiral supervariables. Clearly, under the action of the 
translational generators $((\partial_{\bar{\vartheta}}, \, \partial_\vartheta)$, we get the (anti-)BRST 
invariance of Lagrangian  ($L$) with the  following results  
 \begin{eqnarray}
&& \frac {\partial}{\partial {\bar\vartheta}} \Big[\tilde L ^{(ac)} (t, \bar\vartheta) \Big]  = \frac {d}{d\, t} \Big[{b} (t)\,\dot {c} (t)\Big],\nonumber\\ 
&& \frac {\partial}{\partial {\vartheta}} \Big[\tilde L ^{(c)} (t, \vartheta)  \Big]  \; = \frac {d}{d\,t} \Big[{b} (t)\,{\dot {\bar {c}}}  (t)\Big].
\end{eqnarray}
This implies that the generalized version of the super Lagrangians remains quasi-invariant (i.e., up to a total time derivative) 
under the action of the translational generators $(\partial_{\bar{\vartheta}}, \, \partial_{\vartheta})$ within the framework of ACSA, in a manner consistent with Eq. (10).

\vskip 0.2cm

Next, we examine the (anti-)co-BRST invariance of the Lagrangian (8) within the framework of the (anti-)chiral supervariable approach. 
To achieve this, we extend the standard Lagrangian to the (anti-)co-BRST super Lagrangian, where the (0 +1)-dimensional theory is mapped
onto the (1, 1)-dimensional (anti-)chiral super sub-manifold of the familiar (1, 2)-dimensional supermanifold, as described below:
\begin{eqnarray*}
L(t) & \longrightarrow & \tilde L ^{(c, \,d)} (t, \vartheta) \nonumber\\
& = & P_x ^{(d)} (t, \vartheta)\, \dot X ^{(d)} (t, \vartheta) +  P_y ^{(d)} (t, \vartheta)\, \dot Y  ^{(d)} (t, \vartheta) +  P_z ^{(d)}  (t, \vartheta)\, \dot Z ^{(d)} (t, \vartheta)\nonumber\\
& + & \frac{1}{2}\, \big[P_x ^{(d)} (t, \vartheta)\;P_x ^{(d)} (t, \vartheta) + P_y ^{(d)} (t, \vartheta)\;P_y ^{(d)} (t, \vartheta) + P_z ^{(d)} (t, \vartheta)\;P_z ^{(d)} (t, \vartheta)\big]\nonumber\\
& - & \Xi ^{(d)} (t, \vartheta)\; \big[g \,\{X^{(d)} (t, \vartheta)\; P_y ^{(d)} (t, \bar\vartheta) - Y ^{(d)} (t, \vartheta)\; P_x ^{(d)} (t, \bar\vartheta)\}  +  P_z ^{(d)} (t, \vartheta)\big]\nonumber\\
& + & B^{(d)} (t, \vartheta) \; [\dot \Xi ^{(d)} (t, \vartheta) - Z ^{(d)} (t, \vartheta)]
 + \frac{1}{2}\,B^{(d)} (t, \vartheta)\,B^{(d)} (t, \vartheta)  \nonumber\\
 & - & i\, \dot {\bar F } ^{(d)} (t, \vartheta)\, F^{(d)} (t, \vartheta)  -  i\, {\bar F}^{(d)} (t, \vartheta)\, F^{(d)} (t, \vartheta),
\end{eqnarray*}
\begin{eqnarray}
L(t) & \longrightarrow & \tilde L ^{(ac, \,ad)} (t, \bar\vartheta) \nonumber\\
& = & P_x ^{(ad)} (t, \bar\vartheta)\, \dot X ^{(ad)} (t, \bar\vartheta) +  P_y ^{(ad)} (t, \bar\vartheta)\, \dot Y  ^{(ad)} (t, \bar\vartheta) +  P_z ^{(ad)}  (t, \bar\vartheta)\, \dot Z ^{(ad)} (t, \bar\vartheta)\nonumber\\
& + & \frac{1}{2}\, \big[P_x ^{(ad)} (t, \bar\vartheta)\;P_x ^{(ad)} (t, \bar\vartheta) + P_y ^{(ad)} (t, \bar\vartheta)\;P_y ^{(ad)} (t, \bar\vartheta) + P_z ^{(ad)} (t, \bar\vartheta)\;P_z ^{(ad)} (t, \bar\vartheta)\big]\nonumber\\
& - & \Xi ^{(ad)} (t, \bar\vartheta)\; \big[g \,\{X^{(ad)} (t, \bar\vartheta)\; P_y ^{(ad)} (t, \bar\vartheta) - Y ^{(ad)} (t, \bar\vartheta)\; P_x ^{(ad)} (t, \bar\vartheta)\}  +  P_z ^{(ad)} (t, \bar\vartheta)\big]\nonumber\\
& + & B^{(ad)} (t, \bar\vartheta) \; \big[\dot \Xi ^{(ad)} (t, \bar\vartheta) - Z ^{(ad)} (t, \bar\vartheta)\big] 
 + \frac{1}{2}\,B^{(ad)} (t, \bar\vartheta)\,B^{(ad)} (t, \bar\vartheta)  \nonumber\\
 & - & i\, \dot {\bar F } ^{(ad)} (t, \bar\vartheta)\, F^{(ad)} (t, \bar\vartheta)  -  i\, {\bar F}^{(ad)} (t, \bar\vartheta)\, F^{(ad)} (t, \bar\vartheta).
\end{eqnarray}
Here, the superscripts $(c,\, d)$ and $(ac,\, ad)$ indicate that the super Lagrangians (involving the chiral and {anti-chiral} supervariables) are derived after applying the co-BRST and anti-co-BRST invariant conditions, respectively. It is easy to verify that
\begin{eqnarray}
&& \frac {\partial}{\partial {\vartheta}} \Big[\tilde L ^{(c, d)} (t, \vartheta)  \Big]  \;\; \,= \; -\,\frac {d}{d\, t} 
\Big[p_\varphi  (t)\,\dot {\bar c} (t)\Big],\nonumber\\
&& \frac {\partial}{\partial {\bar\vartheta}} \Big[\tilde L ^{(ac, ad)} (t, \bar\vartheta) \Big]  
= \; -\,\frac {d}{d\,t} \Big[p_\varphi (t)\,\dot c (t)\Big].
\end{eqnarray}
This demonstrates the (anti-)co-BRST invariance of the Lagrangian $L$ within the framework of ACSA to BRST formalism. In conclusion,
this section highlights the following observations: There is a profound relationship between the (anti-)BRST symmetries $(s_{(a)b})$ 
and the derivatives $(\partial_{\bar{\vartheta}}, \; \partial_{\vartheta})$ of the Grassmannian variables $(\bar{\vartheta},\, \vartheta)$, 
represented by the mappings $s_b \longleftrightarrow \partial_{\bar{\vartheta}}$ and $s_{ab} \longleftrightarrow \partial_{\vartheta}$. Similarly,
for (anti-)co-BRST symmetry transformations, these transformations are also related to the derivatives $(\partial_{\bar{\vartheta}}, \; \partial_{\vartheta})$
of the Grassmannian variables, with the mappings $s_d \longleftrightarrow \partial_{\vartheta}$ and $s_{ad} \longleftrightarrow  \partial_{\bar{\vartheta}}$ [see Secs. 4, 5].\\

\section{Conclusions}

In our current analysis, we have derived the off-shell nilpotent quantum (anti-)BRST and (anti-)co-BRST 
symmetry transformations by exploiting standard techniques of the ACSA  to BRST framework where only one of the Grassmannian 
variables (i.e. fermionic variables $\vartheta$ and $\bar\vartheta$ with $\vartheta^2 = 0$ and $\bar\vartheta^2 = 0$) have been used.
The ACSA (a simplified version of the supervariable approach) techniques give the simplest way to derive nilpotent quantum BRST symmetries
because of the presence of only one fermionic variable. Additionally, we have explored the nilpotency as well as absolute anti-commutativity 
properties of the corresponding quantum fermionic (anti-)BRST and (anti-)co-BRST conserved charges for the (0 + 1)-dimensional for gauge-invariant
non-interacting Friedberg-Lee-Pang-Ren (FLPR) model within the framework of ACSA to BRST approach.

\vskip 0.2cm

The key contributions of our present  study include the derivation of off-shell nilpotent (anti-)BRST and (anti-)co-BRST symmetry 
transformations (see Sec. 4), and the demonstration of the nilpotency  ( ${\cal Q}_b ^2 = 0, \; \bar {\cal Q}_{ab} ^2 = 0$) 
and anti-commutativity (${\cal Q}_b\,{\cal Q}_{ab} + {\cal Q}_{ab}\, {\cal Q}_b = 0$) properties of the (anti-)BRST 
as well as  (anti-)co-BRST charges, even though we only considered the (anti-)chiral super expansions of the supervariables onto (1, 1)-dimensional super sub-manifold 
(see Sec. 5). While these properties are naturally expected when using the full super expansions of the supervariables onto (1, 2)-dimensional supermanifold 
(i.e., the BT-supervariable formalism [27-29]), our study successfully demonstrates them using only the (anti-)chiral super 
expansions (with only one Grassmannian variable).

\vskip 0.2cm

It is noteworthy that the nilpotency of the (anti-)BRST conserved charges is linked to the nilpotency 
($\partial^2_{\bar{\vartheta}} = \partial^2_{\vartheta} = 0$) of the translational generators $\partial_{\bar{\vartheta}}$
and $\partial_{\vartheta}$, respectively. Similarly, the nilpotency of the (anti-)co-BRST charges is associated
with the nilpotency ($\partial^2_{\vartheta} = \partial^2_{\bar{\vartheta}} = 0$) of the translational generators 
$\partial_{\vartheta}$ and $\partial_{\bar{\vartheta}}$, respectively. We have also demonstrated (see Sec. 5) that the absolute anti-commutativity of the BRST 
charge with the anti-BRST charge is connected to the nilpotency ($\partial^2_{\vartheta} = 0$) of the translational generator 
$\partial_{\vartheta}$, and the absolute anti-commutativity of the anti-BRST charge with the BRST charge is linked to the nilpotency
($\partial^2_{\bar{\vartheta}} = 0$) of the translational generator $\partial_{\bar{\vartheta}}$. On the other hand, the absolute 
anti-commutativity of the co-BRST charge with the anti-co-BRST charge is tied to the nilpotency of the translational generator
$\partial_{\bar{\vartheta}}$, and the absolute anti-commutativity of the anti-co-BRST charge with the co-BRST charge is deeply related 
to the nilpotency of the translational generator $\partial_{\vartheta}$. 
Moreover, we have demonstrated the present Lagrangian's (anti-)BRST and (anti-) co-BRST invariances within the ACSA framework. 
The action corresponding to the (anti-)chiral super Lagrangian is independent of the Grassmannian variables 
($\vartheta, \bar{\vartheta}$), which is a completely novel result for the present FLPR model (see Sec. 6).

\vskip 0.2cm

In future investigations, as our ACSA standard techniques apply to any theory where gauge invariance (i.e. BRST symmetries) 
is present, we plan to apply the ACSA approach to BRST formalism to various gauge-invariant models and theories such 
as the ABJM theory, supersymmetric Chern-Simons theory, Freedman-Townsend model,
and Abelian gauge theory with higher derivative matter fields.
The idea of ACSA techniques, together with the modified Bonora-Tonin superfield approach (MBTSA), would be useful for discussing various reparameterization invariant models [48-50] particularly for the cosmological Friedmann-Robertson-Walker (FRW) model with a differential gauge condition in the extended phase space [45, 50]. 
Additionally, we will explore the significant and intriguing techniques of ACSA to BRST approach on higher 
$p$-form $(p = 2, 3,...)$ gauge theories in the context of theoretical high-energy  physics from various theoretical
 (e.g. string theory, possible candidates of dark matter and dark energy) and physical perspectives.

\vskip 1.6cm

\noindent
{\bf\large Data Availability}\\

\noindent
No data were used to support this study. \\

\vskip 0.6 cm

\noindent
{\bf\large Conflicts of Interest}\\

\noindent
The authors declare no conflicts of interest that could have influenced the research or its outcomes reported in this paper. All findings are presented objectively and without bias.\\

\vskip 0.9 cm

\noindent
{\bf Acknowledgments:}
One of us (BC) extends appreciation to {\it Sunbeam Women's College Varuna, Varanasi, India} for providing the necessary resources and infrastructure 
to conduct this research. The support and facilities offered were instrumental in completing this work.\\


\begin{thebibliography}{99}
\bibitem{BC1}     C. N. Yang, R. L. Mills,  {\it Phys. Rev.} {\bf 96}, 191 (1954)
\bibitem{BC1}  P. A. M. Dirac, Lectures on Quantum Mechanics, Belfer Graduate\\
School of Science, Yeshiva University Press, New York (1964) 17
\bibitem{BC1}  K. Sundermeyer, Constrained Dynamics: Lecture Notes in Physics,\\
Vol. {\bf 169} Springer-Verlag, Berlin (1982)
\bibitem{BC2}     R. P. Feynman, {\it Acta Phys. Polon.} {\bf 24}, 697 (1963)
\bibitem{BC3}     L. D. Faddeev, V. N. Popov, {\it Phys. Lett.} B {\bf 25}, 29 (1967)
\bibitem{BC4}     B. S. DeWitt, {\it Phys. Rev.} {\bf 160}, 113 (1967); {\bf 162}, 1195 (1967)
\bibitem{RPM5}    C. Becchi, A. Rouet, R. Stora, {\it Phys. Lett.} {\cal B} {\bf 52}, 344 (1974) 
\bibitem{RPM6}    C. Becchi, A. Rouet, R. Stora, {\it Comm. Math. Phys.} {\bf 42}, 127 (1975) 
\bibitem{RPM7}    C. Becchi, A. Rouet, R. Stora, {\it Ann. Phys.} (N. Y.) {\bf 98}, 287 (1976) 
\bibitem{RPM8}    I. V. Tyutin, Lebedev Institute Preprint, Report Number: {\bf FIAN-39} (1975)\\ (unpublished),
                  arXiv:0812.0580 [hep-th]
\bibitem{BC9}     I. A. Batalin, P. M. Lavrov, I. V. Tyutin, {\it J. Math. Phys.} {\bf 31}, 1487 (1990)
\bibitem{BC10}    I. A. Batalin, P. M. Lavrov, I. V. Tyutin, {\it J. Math. Phys.} {\bf 32}, 532 (1991)
\bibitem{BC11}    I. A. Batalin, P. M. Lavrov, I. V. Tyutin, {\it J. Math. Phys.} {\bf 32}, 2513 (1991)
\bibitem{BC13}    R. Friedberg, T. D. Lee, Y. Pang,  H. C. Ren, {\it Phys. Rev.}  D, {\bf 46}, 4119 (1992)
\bibitem {BC14}   R. Friedberg, T. D. Lee, Y. Pang,  H. C. Ren, {\it Ann. Phys.} {\bf 246}, 381 (1996)
\bibitem {BC15}   V. N. Gribov, {\it Nucl. Phys.} B {\bf 139}, 1 (1978)
\bibitem {BC16}  R. Banerjee, arXiv:hep-th/9610240
\bibitem {BC17}  V. M. Villanueva, J. Govaerts, and J. -L. Lucio-Martinez,
                  {\it J. Phys.} A {\bf 33}, 4183 (2000)
\bibitem {BC18}   R.  Friedberg,  T. D. Lee, Y.. Pang,  J. Ren,  {\it Phys. Review } D, {\bf 40}, 1147 (1989)
\bibitem {BC19}   K. Fujikawa, {\it Nucl. Phys.} B {\bf 468}, 355 (1996)
\bibitem{BC20}  R. P. Malik, {\it Euro. Phys. Lett.} {\bf  144}, 42002 (2023) 
\bibitem{BC21}  S. Krishna, R. P. Malik, {\it Ann. Phys.} (N. Y.) {\bf 464}, 169657 (2024) 
\bibitem{BC22}  A. S. Nair, R. Kumar, S. Gupta, {\it Eur. Phys. J. Plus} {\bf 138}, 1107 (2023)
\bibitem{BC23}  A. S. Nair, S. Gupta, {\it Mod. Phys. Lett. } A {\bf 39}, 2350186  (2024) 
\bibitem {BC24}   J. Thierry-Mieg, {\it J. Math. Phys.} {\bf 21}, 2834 (1980)
\bibitem {BC25}   M. Quiros, F. J. De Urries, J. Hoyos, M. L. Mazon, E. Rodrigues,\\ 
                  {\it J. Math. Phys.} {\bf 22}, 1767 (1981)
\bibitem {BC26}   L. Bonora, M. Tonin, {\it Phys. Lett.} B {\bf 98}, 48 (1981)
\bibitem {BC27}   L. Bonora, P. Pasti, M. Tonin, {\it Nuovo Cimento} A {\bf 64}, 307 (1981)
\bibitem {BC28}   L. Bonora, P. Pasti, M. Tonin, {\it Ann. Phys} {\bf 144}, 15 (1982)
\bibitem {BC29}   R. P. Malik, {\it J. Phys. A: Math. Theor.} {\bf 39}, 10575 (2006)
\bibitem {BC30}   R. P. Malik, {\it Eur. Phys. J.} C {\bf 51}, 169 (2007)
\bibitem {BC31}   R. P. Malik, {\it Eur. Phy. J.} C {\bf 60}, 457 (2009)
\bibitem {BC32}   I. A. Batalin, K. Bering, P. H. Damgaard, {\it Nucl. Phys.} B {\bf 515}, 455 (1998)
\bibitem {BC33}   I. A. Batalin, K. Bering, P. H. Damgaard, {\it Phys. Lett.} B {\bf 446}, 175 (1999)
\bibitem {BC34}   P. M. Lavrov, P. Yu. Moshin, A. A. Reshetnyak,\\
                   {\it Mod. Phys. Lett.}  A {\bf 10}, 2687 (1995); {\it JETP Lett.} {\bf 62}, 780 (1995)
\bibitem {BC36}   N. Srinivas, T. Bhanja, R. P. Malik, {\it Adv. High Energy Phys.}  {\bf 2017}, 6138263 (2017)

 \bibitem {BC37}    S. Kumar, B. Chauhan, R. P. Malik {\it Int. J. Mod. Phys.} A {\bf 33},  1850133 (2018)
\bibitem {BC37}  B. Chauhan, S. Kumar, R. P. Malik, {\it  Int. J. Mod. Phys.} A {\bf 33}, 1850026 (2018)
\bibitem {BC38}  A. Shukla, N. Srinivas, R. P. Malik, {\it Ann. Phys.} {\bf 394}, 98 (2018)
\bibitem {BC37}   T. Bhanja, N. Srinivas, R. P. Malik, 	{\it Int. J. Mod. Phys.}  A {\bf 34},  1950183 (2019) 
 \bibitem {BC37}   B. Chauhan, S. Kumar, R. P. Malik,  {\it Int. J. Mod. Phys.} A {\bf 34}, 1950131 (2019)
\bibitem {BC37}   B. Chauhan, S. Kumar, A. Tripathi, R. P. Malik, \\ {\it Adv. in High Energy Phys.} {\bf 2020}, 3495168, (2020)
\bibitem {BC35}  B. Chauhan, S. Kumar, {\it Adv. High Energy Phys.}  {\bf 2021}, 5518304  (2021)
 \bibitem {BC35}    S. Kumar, B. Chauhan, A. Tripathi, R. P. Malik, \\
   {\it Int. J. Mod. Phys.} A {\bf 37},   2250003 (2022)
\bibitem {BC36}  B. Chauhan, {\it Eur. Phys. J. Plus} {\bf 137}, 976 (2022) 
\bibitem {BC37}  B. Chauhan,  {\it Eur. Phys. Lett.} {\bf 140}, 40001 (2022) 

\bibitem {BC39}  B. Chauhan, S. Kumar, R. P. Malik, {\it Int. J. Mod. Phys.} A {\bf 37}, 2250003 (2022)
\bibitem {BC40}  B. Chauhan, A. Tripathi, A. K. Rao, R. P. Malik, 
                    \\ {\it Int. J. Mod. Phys. } A {\bf 37} 2250164 (2022) 
\bibitem {BC41}  A. Tripathi, B. Chauhan, A. K. Rao, R. P. Malik,\\
                    {\it Adv. High Energy Phys.}  {\bf 2021}, 2056629 (2021)
\bibitem {BC41}  B. Cahuhan, R. Tripathi, {\it Nucl. Phys.} B {\bf 1012}, 116816 (2025)                    












\end{thebibliography}
\end{document}